\documentclass[prb,preprint,epsfig,floats]{revtex4}
\usepackage{graphicx,epsf}

\begin{document}


\title{ A Theory of the Pseudogap State of the Cuprates}
\author{C. M. Varma}
\address{Physics Department, University of California,
Riverside, CA 92507}
\begin{abstract}

The phase diagram for a general model for Cuprates is derived in a mean-field approximation. A phase violating time-reversal without breaking translational symmetry is  possible when both the ionic interactions and the local repulsions are large compared to the energy difference between the Cu and O single-particle levels. It ends at a quantum critical point as the hole or electron doping is increased. Such a phase is necessarily accompanied by singular forward scattering such that, in the stable phase, the density of states at the chemical potential, projected to  a particular point group symmetry of the lattice is zero producing thereby an anisotropic gap in the single-particle spectrum. It is suggested that this phase occupies the "pseudogap" region of the phase diagram of the cuprates. The temperature dependence of the single-particle spectra, the density of states, the specific heat and the magnetic susceptibility are calculated with rather remarkable correspondence with the experimental results. The importance of further direct experimental verification of such a phase in resolving the principal issues in the theory of the
Cuprate phenomena is pointed out. To this end, some predictions are provided.
\end{abstract}
\maketitle

 \section{Introduction}

It became quickly apparent following the initial discovery
\cite{BM} that the high superconducting transition temperatures
   in the Cuprates  was not the only
    interesting feature of these compounds. The properties
   in the normal state(s) are unlike other metals and  new concepts
   are required to understand them. Moreover the superconductivity can be understood
   only if they are understood since the fluctuations responsible for the
   superconducting instability are the property of the normal state.

 The extensive investigation of the Cuprates \cite{expts},\cite{timusk}
has led to a consistent
 set of experimental results from which a universal phase diagram, fig. (1), for these
  compounds was drawn \cite{cmv1,cmv2}, and for which considerable further evidence has been adduced \cite{tallon, panagoupoulous,alff, timusk, greene}
  This diagram was drawn on the basis of the
  measured properties,
  thermodynamic as well as transport,which
  show characterestic changes across the lines drawn {\it in all the cuprates}.
  A fundamental aspect of the phase diagram is the existence of a {\it putative} quantum Critical
  point (QCP) inside the regime of compositions for superconductivity.
Besides the  superconducting region, there lie, as marked in fig. (1),  three distinct regions emanating from the QCP. This paper is concerned primarily with region (II) of the
phase diagram, but as explained below, its nature is tied to the nature of the QCP and the fluctuations above it and understanding it amounts to understanding the essentials of the  physics  of the Cuprate phenomena.

     The QCP was anticipated by the {\it marginal fermi-liquid hypothesis}\cite{cmv3}, which proposed a scale invariant spectrum of fluctuations to explain the normal state properties in region (I) and led to a set of predictions for the single-particle spectra and transport properties which have been verified \cite{arpes} \cite{kaminski-pr}. The suggestion\cite{cmv3} that the same spectrum provides a glue for superconductive pairing is also consistent with that deduced from approximate inversion of angle-resolved photoemission (ARPES) data \cite{norman} in the superconductive state. The proposed spectrum has the form
     \begin{eqnarray}
     Im\chi({\bf q}, \omega, T) & \propto & \omega/T, ~ for~ \omega \ll T,   \\  \nonumber
                                              & \propto & constant, ~ for ~ T \ll \omega \ll \omega_c,
    \end{eqnarray}
  where  $\omega_c$ is a cut-off, determinable from experiments. The corresponding real part of the fluctuations is $Re\chi({\bf q}, \omega, T) \propto \ln(\omega_c/\omega)$
  for $\omega/T \gg 1$ and $\propto \ln(\omega_c/T)$ for $\omega/T \ll 1$.

  \begin{figure}[htbp]
\begin{center}
\includegraphics[width=0.8\textwidth]{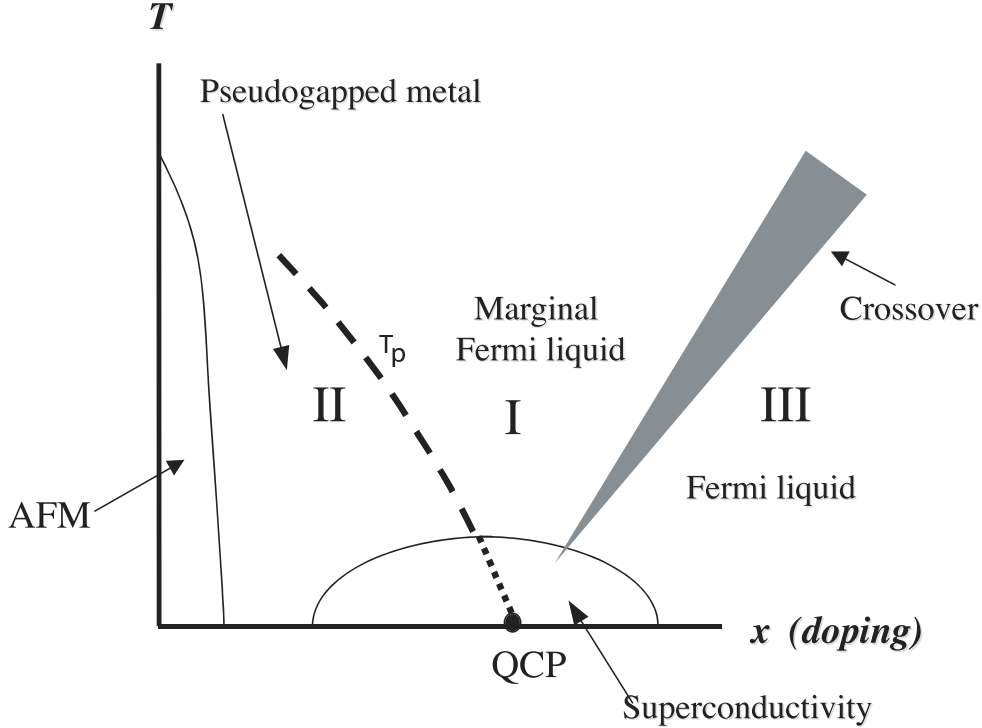}
\caption{A schematic of the {\it universal} Phase diagram for the Cuprates; the lines shown and the crossover occur in all the Cuprate compounds. Other features, small regions of antiferromagnetism away from half-filling, charge density modulations, stripes, etc. which occur in one or the other Cuprates as well as the spin-glass region are not shown.}
\label{fig. 1}
\end{center}
\end{figure}

   It is characterestic of QCP's in itinerant fermions \cite{sachdev, cmvleiden} that a crossover to  a fermi-liquid is expected characterized by a line emanating from $x=x_c$. Below another line emanating from $x=x_c$, a broken symmetry is expected. These expectations are consistent with the experimental phase diagram of fig. (1): characterestic Fermi-liquid properties are found deep in region (III) and in region (II) an anisotropic gap (pseudogap) in the one-particle spectra at the chemical potential is observed \cite{arpespseudogap}. Transport \cite{panagoupoulous} and thermodynamic properties \cite{susc, spheat} also show characteristic changes below the line $T_p(x)$. However the nature of region (II) remains in doubt. There is plenty of empirical evidence to conclude that region (II) does not universally break translational and/or spin-rotational symmetries. If it is a phase of broken symmetry, it is of an unusual nature.

   A knowledge of the ground state and excitations of region (II) ties together every essential aspect of the Cuprate problem. Given a QCP, region (III) follows automatically.  Given the success of the predictions following from the phenomenological spectrum of Eq.(1), one can be reasonably certain about the functional form of the spectrum. Since the broken symmetry in Region (II) is due to the condensation of the fluctuations in region (I), a knowledge of the former specifies the nature of the latter. It answers the question, the fluctuations of which operator have the spectra given by Eq. (1)? There is experimental evidence \cite{norman} that Eq.(1)  also gives the glue for the pairing instability for superconductivity. Therefore a knowledge of region (II) specifies the pairing mechanism as well.

  A hint is available about the possible broken symmetry in Region (II)  from the Raman spectra, which measures the $ q \rightarrow 0$ part of the particle-hole fluctuations in even parity point-group symmetries. In region (I), it is firmly established that the spectrum is consistent \cite{slakey} with the predicted form, Eq. (1).  For conserved quantities like the density or spin-density, the correlation functions must be proportional to $q^2$ at small $q$. Therefore one infers that the observed singularity is in the correlation function of an un-conserved operator in some irreducible representation (IR) of the lattice at
        $ q=0$. (This IR is inferred from Raman experiments to be the $B_{1g}$ representation.) To be consistent with the existence of a QCP, the lowered symmetry in region (II) must then be due to the condensation of such an operator. In the ordered state, the singularity in the Raman response must then disappear; this is also consistent with experiments in region (II) \cite{cardona}.

        An operator satisfying the above requirements is the {\it current} operator at $q=0$ in the $B_{1g}$  representation of the lattice \cite{footnote1},\cite{footnote 2}. In a microscopic theory of the cuprates \cite{cmv1, cmv2} based on a model suggested earlier
     \cite {giam},
    it was predicted that the region (II) violates time-reversal symmetry
    by  a ground state which carries currents in the 0-Cu-0 plaquettes in each
    cell.
    An experiment using ARPES with circularly
    polarized photons was
    suggested \cite{cmv4} to identify the T-breaking phase
    phase. Some detailed experiments have borne out these predictions \cite{kaminski}.
    New classes of experiments have been suggested to further verify
     the predicted phase \cite{dimatteo, ng}.

     The violation of time-reversal symmetry is shown here to be accompanied by
      singular attractive forward scattering due to the generation of a retarded $1/q^2$ interaction. Due to such a scattering, the density of states of one-particle spectra at the chemical potential, projected to a particular (IR)  of the lattice tends to zero.  The resulting stable state is shown to have an anisotropic gap in the single-particle spectra at the chemical potential \cite{cmv2}.

     It is worthwhile summarizing  the principal results observed in region II of the phase diagram. The principal property in this region is an anisotropic loss of low energy excitations. First discovered in measurements of Knight shift\cite{susc}, this was soon found in the specific heat measurements\cite{spheat}, as well as in direct measurements of the single-particle spectra through ARPES\cite{arpespseudogap}. Transport properties, Raman scattering and neutron scattering  also show characterestic changes in their temperature (frequency) dependences below  $~T_p(x)$. This region is often said to have a {\it pseudogap}. This is a misnomer, (which I am forced to persist in using.) ARPES measurements show
     (see for example, figs. (26) and (28) of Ref. (\onlinecite{campu-rev})) that the anisotropic gap is as robust as the superconducting gap at a corresponding reduced temperature.

The principal purpose of this paper is to give  details of the results presented earlier \cite{cmv2,errors} and to address several related new questions.
 I  focus here
     on the broken symmetry as well as the generated anisotropic gap and the  equilibrium
    properties and the single-particle spectra in region (II).  The density of states, the specific heat and the magnetic susceptibility are calculated with remarkable correspondence with the experiments. The
     transport properties will be studied in the near future. The paper is
    organized as follows: The motivations for the specific model introduced for the Cuprates are reiterated in Sec.II,  where the physics of time-reversal violation or local circulating currents is related to the screening of charge fluctuations special  to  the cuprates.
     Time-reversal violating (TRV) states
    are derived in mean-field calculations in Sec.III.
      A phase
     diagram similar to fig. (1) is derived.
 The current patterns and the symmetries in this
    state are also discussed. In Sec.IV, the coupling functions
    of the collective fluctuations towards the TRV states to the
    fermions are derived to show why time-reversal violation must necessarily engender a fermi-surface instability. The stable state is shown to have an anisotropic gap at the chemical potential.
     The single-particle spectrum and the
    thermodynamics are derived and compared with experiments and some predictions for further experiments may be found in Sec. VI.

     An important new result is that the derived coupling of the fluctuations in Region (I) to the fermions suggests that the phenomenological spectrum of Eq.(1) should be modified in one respect: $\chi({\bf q}, \omega, T)$ should be replaced by $\chi({\bf q},{\bf k}, \omega, T)$ where at small momentum transfer ${\bf q}$, the ${\bf k}$ dependence near the fermi-surface varies with the dominant IR of the fluctuations.

  Various sundry
    issues, including the limitations of the theory are discussed in the concluding section.

\section{The Minimal Model for Cuprates}

A microscopic model for the Cuprates must take into account that the phenomena observed in them hardly ever, if at all, occurs in any other known compound. The effective Hamiltonian for the Cuprates must therefore reflect some unique features of their solid-state chemistry.  Consideration of this issue led to the 3-orbital
 model with
longer-range interactions as well as  the local repulsions on Cu and on
O \cite{giam}.

 Cuprates  have the unique  feature that the ionization energy of the $Cu^{++} $ state,  is nearly equal to  the affinity energy of the $O^{--}$  state, i.e.  $ E(Cu^{+})-E(Cu^{++})$ is nearly equal to  $E(O^-)-E(O^{--})$ \cite{zsa}. The corresponding difference is much larger than the one-particle transfer integrals in other transition metal oxides.  Therefore  the charge fluctuations in the metallic state occur almost equally on Copper and on Oxygen ions.   The 4s-p orbitals on Cu are too far away in energy to screen the mixed-valence fluctuations on Cu intra-atomically; the 3s state of O are similarly too far away for intra-atomic screening. (These orbitals, when they control the screening, can be eliminated in favor of a renormalized effective repulsion parameter, the U of the Hubbard-model.) Charge fluctuations can then only be screened by complementary charge fluctuations on the neighboring ions. A consequence is that the screening length is close to the nearest neighbor Cu-O distance giving large effective ionic interactions which summed over the nearest neighbors are of the same order as the local repulsion energies (U) on the Cu-or O ions. Since both the ionization-affinity energy difference and the transfer integrals are smaller than the ionic interactions, there is  no small parameter in the physics of the  Cuprates by which the ionic interactions may be absorbed to start with a simpler model such as the Hubbard and $t-J$ model proposed \cite{anderson} for the Cuprates.

 Whether the ionic interactions introduce qualitatively new properties can only be decided by calculations using them. I show in this paper that this is possible using approximate but systematic methods. For a range of ionic interactions and doping,  screening can be achieved coherently through occupied wave-functions which are complex admixtures of orthogonal orbitals in a unit-cell, i.e.  by currents in the elementary O-Cu-O plaquettes.  Such coherent screening represents time-reversal violation.

 The minimal model Hamiltonian \cite{giam} is
on the basis
of three orbitals per unit cell, $d,p_x,p_y$:
\begin{equation}
H = K + H^{(1)}_{int} + H^{(2)}_{int}.
\end{equation}
$K$ is the kinetic energy operator,
\begin{equation}
K = \sum_{{\bf k}, \sigma} \epsilon_d n_{d {\bf k} \sigma} + 2  \: t_{pd}
d^+_{{\bf k}, \sigma} \left( s_x (k) p_{x {\bf k} \sigma} +
s_y (k) p_{y {\bf k} \sigma}\right) - \: 4 t_{pp} \:
s_x (k) s_y (k) p^+_{x {\bf k} \sigma} p_{y {\bf k} \sigma} + H.C.
\end{equation}
Here a particular choice of the relative phases of
the x and y orbitals in the unit cell has been made,
$s_{x,y}(k) = sin (k_x a/2 , k_ya / 2 )$ and for later,
$c_{x,y}(k)= cos(k_x a/2 , k_y a/2 )$ and
$s_{xy}^2(k) = (sin^2(k_x a/2) + sin^2 (k_y a/2 ))$.
The local interaction on the Cu and the O orbitals are
\begin{equation}
H^{(1)}_{int} = \sum_{i, \sigma} U_d \: n_{di \sigma}  \:
n_{di - \sigma} + U_p \left( n_{p x i \sigma} \;
n_{pxi - \sigma} + \; n _{pyi \sigma} n_{pyi - \sigma} \right)
\end{equation}
and the nearest neighbor interaction between the Cu and the O orbitals are
\begin{eqnarray}
H^{(2)}_{int}= \sum_{i,nn} V n_{di}(n_{i+nn,p_x} +n_{i+nn,p_y}),
\end{eqnarray}
where $(i+nn)$ stands for nearest O neighbors of $Cu$ in a cell $i$ with $(p_x,p_y)$ orbitals in the $(x,y)$ direction.
In momentum space, this is
\begin{equation}
H^{(2)}_{int} = 2V \sum_{{\bf k} {\bf k}^{\prime} {\bf q}, \sigma \sigma^{\prime}}
c_x (q) \: d^+_{{\bf k+q} \sigma} \: d_{{\bf k} \sigma}  \:
p^+_{x{\bf k-q}\sigma'}
p_{x {\bf k}^{\prime} \sigma^{\prime}} + x \rightarrow y
\end{equation}

We may also include the nearest neighbor interactions between the
O orbitals or longer range interactions. Such interactions do not change the essential results
derived here provided they are small compared to $V$. In the calculations in this paper (renormalised)
energy difference $\epsilon_d$ between the Cu and the O orbitals
is taken zero. It is important that in CuO, $\epsilon_d \lesssim
O(t_{pd})$. Taking it as zero simplifies the calculations
presented; the principal effect of a finite $\epsilon_d$ will be
mentioned.

In the limit that $(U_d, U_p)\gg (t_{pd}, t_{pp})$, a good
approximation \cite{constraint} for small deviations $|x|$
from half-filling consists in replacing \cite{constraint}
\begin{eqnarray}
t_{pd} \rightarrow \bar{t}_{pd} = t_{pd}|x|;~~
 t_{pp} \rightarrow \bar{t}_{pp} = t_{pp}|x|,
 \end{eqnarray}
 where $|x|$ os the deviation from half-filling; $x > 0$ for holes and
 $x < 0$ for electrons. The Lagrange parameter  appropriate to this transformation
 is absorbed in the chemical potential. A more general and messy mean-field calculation,
  would consider separately
  the average occupations in the oxygen and copper orbitals and renormalize
  $t_{pd}, t_{pp}$ accordingly. It is not expected to change any
  of the essential results.

  An alternate treatment \cite{cmv1} to handle the large local and ionic
  repulsions  is to start from the strong-coupling limit by diagonalizing a unit-cell exactly \cite{sire, perakis} by cutting of its kinetic energy connection to its neighbors and then to introduce the kinetic energy connecting the lowest energy many body states in each cell. This is the generalization of the procedure by which the $t-J$ hamiltonian is derived from the Hubbard model. Starting from the model of Eq. (1), this leads minimally to  an effective $SU(2)\times SU(2)$ hamiltonian plus the constrained kinetic energy; one $SU(2)$ represents the usual spin-space while the other represents the two lowest many-body states obtained in each cell by the first step in the procedure.
  \begin{equation}
  H_{int} = {\cal J}(-1/2+\sigma_i\cdot\sigma_j)({\bf A} + {\bf \tau}_i{\bf B}{\bf \tau}_j). \nonumber
  \end{equation}
  Here ${\bf \tau}$'s are in  the orbital space and ${\bf A,B}$ give the anisotropy of interactions in that space.
  This hamiltonian leads to solutions with the same symmetries as in the approach below.

  \section{Derivation of Time-Reversal Violating States}

Time-reversal violating (TRV) states are shown here in a mean-field approximation  to be possible stable states for the model in a range parameters.
To construct a mean-field theory,  write the interactions of Eq.(5) in a separable form.
\begin{eqnarray}
 H^{(2)}_{int}  & = & -V\sum_{i=1..4,{\bf k,q},\sigma}
 A^{\dagger (i)}_{{\bf q,k},\sigma}
 \sum_{{\bf k'},\sigma'} A^{(i)}_{{\bf k',q},\sigma'}.   \\
 A^{(1,2)}_{{\bf k,q},\sigma} & = & \left(
 s_x({\bf{k}})p_{x,{\bf{k+q}},\sigma}^{\dagger} d_{{\bf{k}},\sigma}
 \pm s_y({\bf{k}})p_{y,{\bf{k+q},\sigma}}^{\dagger} d_{{\bf{k}},\sigma} \right) \\
 A^{(3,4)}_{{\bf k,q},\sigma} & = &  \left(
 c_x({\bf{k}})p_{x,{\bf{k+q}},\sigma}^{\dagger} d_{{\bf{k}},\sigma}
 \pm c_y({\bf{k}})p_{y,{\bf{k+q}},\sigma}^{\dagger} d_{{\bf{k}},\sigma}\right).
 \end{eqnarray}
 The superscripts $(1,3)$ and $(2,4)$ refer to the plus
 and minus signs respectively.

A model
 such as that in Eq. (2) is a severe simplification of the
 problem. It may be the minimum essential model but the parameters
 that faithfully represent the actual problem through it are
 knowable to factors of 2 at best.  Staggered flux phases \cite{affleck, chakravarty}, charge-density
 waves \cite{castellani1}, spin-density waves, charge-transfer instabilities \cite{castellani, littlewood} etc.
 may be produced in mean-field by the same Hamiltonian by varying different parameters.
It is not the purpose of this paper to find the phase diagram for
 all these instabilities in the space of the parameters of the
 Hamiltonian, but to show that the trial states chosen for investigation
 by discarding those that can be excluded on the basis of experiments occur for reasonable
 physical parameters, and let
 experiments decide whether they indeed occur.

Accordingly, I choose to look only for mean-field solutions which do not break translational symmetry
 and/or spin-rotational symmetry. For such solutions, one needs to consider only the zero-momentum transfer, $q=0$, part of the Hamiltonian
 $H^{(2)}_{int}$. We can therefore introduce possible mean-field order parameters $R_{i}exp(i\phi_{i})$ and rewrite
 \begin{eqnarray}
H^{(2)}_{int}  =  -
V\sum_{{\bf q},i}[\sum_{{\bf k},\sigma}A^{(i)\dagger}_{{\bf{k,q}},\sigma} - \delta_{{\bf q},0}R_{i}/V \exp(-i\phi_{i})]
    [\sum_{{\bf{k'}},\sigma}
     A^{(i)}_{{{\bf k',q}},\sigma} -
\delta_{{\bf q},0}R_{i}/V \exp(i\phi_{i})]   \nonumber  \\
+  \sum_i R_{i}^2/V - R_{i} \sum_{{\bf{k}},\sigma}[A^{(i)}_{{{\bf {k},0},\sigma}}
\exp(i\phi_{i}) + A^{(i)\dagger}_{{{\bf{k,0}},\sigma}}
\exp(-i\phi_{i})],    \nonumber  \\
R_{i}exp(i\phi_{i})  =  V \sum'_{{{\bf k}},\sigma}\langle A^{(i)}_{{{\bf k},0},\sigma} \rangle .
\end{eqnarray}
 The sum over ${\bf k}$ in $R$'s is restricted to the occupied part and the expectation value is in the state to be determined self-consistently.

 The mean-field Hamiltonian
 with order parameter at ${\bf q} =0$  can only be a
 $3\times 3$ matrix in the space of the three orbitals per unit cell at each ${\bf k}$.
 Furthermore, the
  effective single particle energy for the $p_x$ and the $p_y$ orbitals must
  be kept degenerate; the renormalization of their difference from the energy
   of the d-orbital can produce a
  charge transfer instability \cite{littlewood} which is generically of first order and therefore
   not of interest. ( We will take the point of view that such renormalizations
   have already been taken into account in the bare parameters of the theory.)
   This leaves only three functions of ${\bf k}$ in the off-diagonals  to be determined.
The $q=0$ part of $ A^{(1)}$ transforms simply as the
existing kinetic energy
 and cannot lead to a broken symmetry. If a real order parameter is constructed from
 the rest of $ H^{(2)}_{int}$, a structural distortion of the unit cell is a
 necessary accompaniment. We discard this possibility as well. So the order parameter must be
 non-real. As will be obvious below,
$A^{(3)}$ and $ A^{(4)}$ lead to the different domains of the same phase.
 So a mean-field theory of $H^{(2)}_{int}$  of Eq. (5), not in conflict with experiments, can produce only two distinct order parameters:
 \begin{eqnarray}
     R_I \exp (i\phi_I)  & = & V \sum'_{{{\bf k}},\sigma}\langle A^{(2)}_{{{\bf k},0},\sigma} \rangle = \pm V\sum'_{{\bf k},\sigma}
 [s_x \langle p^{+}_{x{\bf {k},\sigma}}
 d_{\bf {k},\sigma}\rangle
    - s_y \langle p^{+}_{y,{\bf k},\sigma}d_{\bf k}\rangle].  \\
 R_{II} \exp (i\phi_{II})  & = & V \sum'_{{{\bf k}},\sigma}\langle A^{(3,4)}_{{{\bf k},0},\sigma} \rangle = \pm V\sum'_{\bf {k},\sigma}
 [c_x \langle p^{+}_{x{\bf {k},\sigma}}
 d_{\bf {k},\sigma}\rangle
    \pm c_y \langle p^{+}_{y,{\bf {k},\sigma}}d_{\bf {k},\sigma}\rangle].
 \end{eqnarray}

In the mean-field approximation the expectation value of the constrained kinetic energy plus $H^{(2)}_{int}$ is minimized to determine the order parameter $R$.    For the state $\Theta_{I}$, the mean-field
hamiltonian is

\begin{equation}
{\cal H}^I_{mf} = K+ R_I^2/V- R_1\exp({-i\phi}_I)\sum_{{\bf{k}},\sigma}
[s_x({\bf k})p^{+}_{x{\bf k},\sigma}
 d_{{{\bf k}},\sigma}-s_y({\bf k}) p^{+}_{y{\bf k},\sigma}
 d_{{\bf {k}},\sigma}] + H.C.
 \end{equation}

 For the state $\Theta_{II}$, the mean-field Hamiltonian is

 \begin{eqnarray}
{\cal H}^{II}_{mf} = K+ R_{II}^2/V- R_{II}\exp({-i\phi}_{II})\sum_{{\bf{k}},\sigma}
[c_x({\bf k})p^{+}_{x{\bf k},\sigma}
 d_{{{\bf k}},\sigma}-c_y({\bf k}) p^{+}_{y{\bf k},\sigma}
 d_{{\bf {k}},\sigma}] + H.C.
 \end{eqnarray}

 For both states the first and the third terms of the {\it  mean-field} Hamiltonian written in the space
 of the orbitals $ d_{\bf k},p_{x {\bf k}},p_{y {\bf k}}$ is the
  matrix
  \begin{eqnarray}
 {\cal H}_{mf} =
 \left(
 \begin{array}{lcr}
                                           0       & 2\bar{t}_{pd}e^{i\theta_x(\bf{k})}
                                                                 s_x(\bf{k})      &
                                                                                             2\bar{t}_{pd}e^{i\theta_y(\bf{k})}s_y(\bf{k}) \\
                      2\bar{t}_{pd}e^{-i\theta_x(\bf{k})}s_x(\bf{k})
                                                   &                   0               & 4\bar{t}_{pp}s_x(\bf{k})s_y(\bf{k})  \\
   2\bar{t}_{pd}e^{-i\theta_y(\bf{k})}s_y(\bf{k})
                                                   &  4\bar{t}_{pp}s_x(\bf{k})s_y(\bf{k})
                                                   &                       0
  \end{array}
   \right)
  \end{eqnarray}

  For the state $\Theta_I$
 \begin{eqnarray}
\tan\theta_x = \frac{R_I\sin\phi_I}{2t_{pd}+R_I\cos\phi_I}, \\ \nonumber
\tan\theta_y = \frac{-R_I\sin\phi_I}{2t_{pd}-R_I\cos\phi_I}
  \end{eqnarray}
while for the state $\Theta_{II}$
\begin{eqnarray}
\tan(\theta_{x}(k))= \frac{R_{II}c_x
\sin\phi_{II}}{2\bar{t}_{pd}s_x+R_{II}c_x\cos\phi_{II}}
\\ \nonumber
\tan(\theta_{y}(k))=\pm \frac{R_{II}c_y
\sin\phi_{II}}{2\bar{t}_{pd}s_y\pm R_{II}c_y\cos\phi_{II}}.
\end{eqnarray}

   The unitary (gauge) transformation
\begin{eqnarray}
d_{\bf{k},\sigma} \rightarrow d_{\bf{k},\sigma}; ~~ p_{x,\bf{k},\sigma}
 \rightarrow p_{x,\bf{k},\sigma}\exp(i\theta_x); ~~p_{y,\bf{k},\sigma}
 \rightarrow p_{y,\bf{k},\sigma}\exp(i\theta_y) \nonumber
 \end{eqnarray}
 on Eq. (17), transforms it to
\begin{eqnarray}
\left( \begin{array}{clcr}
                                           0 &
                                           2\bar{t}_{pd}s_x(\bf{k})

                                                                                & 2\bar{t}_{pd}s_y(\bf{k}) \\
   2\bar{t}_{pd}s_x(\bf{k})  &              0               & 4\bar{t}_{pp} e^{i(\theta_x-\theta_y)}s_x({\bf k})s_y({\bf k})  \\
   2\bar{t}_{pd}s_y(\bf{k})  &  4\bar{t}_{pp}e^{i(\theta_y-\theta_x)}s_x({\bf k})s_y({\bf k})
                                                                               &                   0
  \end{array} \right).
   \end{eqnarray}

   Note that
\begin{equation}
\theta({\bf k}) \equiv (\theta_x({\bf k})-\theta_y({\bf k})),
\end{equation}
 is the invariant {\it flux} associated with the
 elementary O-Cu-O  plaquette. (The correspondence of the theory  here to the generation of time-reversal violation in $SU(3)$ model of elementary particles through the Cabibo, Kobayashi, Masakawa $SU(3)$ mass matrix \cite{hepbook} may be noted. The time-reversal violation there is in an abstract space; here  the phenomena is more vividly pictured.) The problem therefore reduces to determining this invariant by minimization of the Free-energy.

Returning to the minimzation,
 since at any $\bf{k}$ the trace over the three bands is conserved, the change in ground state
 energy can be calculated from the change in energy of the unfilled part of the
 conduction band alone. In  an Appendix these are calculated to leading order in $t_{pp}/t_{pd}$. The
 value of the phase $\phi_{I,II}$ at the minimum is found to be at
 \begin{eqnarray}
 \phi = \pm \pi/2.
 \end{eqnarray}
 With $\epsilon_d =0$, and the deviation from half-filling $|x| \ll 1$,
 $\theta_{1,2} \ne 0$ at $T=0$, only if
 \begin{equation}
 V\gtrsim 2a_{1,2} \frac{|x|}{(1+x)}t_{pd}^3/t_{pp}^2,
 \end{equation}
where  $a_{1,2}$ are numerical coefficient of order unity, which depend
  on the details of the band-structure and whether the state is $\Theta_I$ or $\Theta_{II}$. The equality in Eq.(23) provides the hole ($x_c>0$) and the electron ($x_c<0$) densities at the QCP's.   In Appendix A it is shown that the
  simplest model favors the $\Theta_{II}$ phase. Only the  possibility of the $\Theta_I$ phase was considered earlier \cite{cmv2}. (23) gives that for $V \simeq t_{pd} \simeq 2t_{pp}$, $x_c \simeq 0.1$ which, in view of the approximations made, is surely a stroke of luck.

 In the Appendix the transition temperature is also derived to find
  \begin{eqnarray}
  T_g \propto E_f/ln(1-x/x_c),~~x \le x_c,
  \end{eqnarray}
 where $E_f$ is the Fermi-energy in the conduction band. At $x\rightarrow 0$, this
  result is an artifact of the model with $\epsilon_d$ taken zero.
  (At $x\rightarrow 0$, the effective masses in all bands are
  infinite, so that for $\epsilon_d =0$ the bands are degenerate and  the transition occurs for
  $V\rightarrow 0$ at $T\rightarrow 0$ or infinite $T$ at $V$
  finite.) Moreover competition of antiferomagnetism and the TRV
  state has not been considered here. The result is therefore only valid for
  $x$ close to $x_c \ll 1$. Eqs. (23,24) give the  correct general
  shape for the phase (II) marked in fig. (1) for $x$ below near $x_c$.

The mean-field wave-function is made up of products of
 $|\bf{k},\theta_1\sigma\rangle$ up to the Fermi-vector. In the
 original gauge, i.e. in which Eqs. (2) and (15) are written, the
 new (TRV) conduction band wave-functions for either state $\Theta_I$ or $\Theta_I$ to leading order in $t_{pp}/t_{pd}$ are, ( the spin-index is suppressed),
 \begin{eqnarray}
 c^{+}_{{\bf k},\theta_1}|0 \rangle   =  (N_k)^{-1}[a^+_{{\bf k}} +
2\frac{\bar{t}_{pp}s_xs_y}{\epsilon^0_{\bf{k}}s_{xy}^2}
 \left( (s_x^2-s_y^2)\cos(\theta_{(1,2)}({\bf k}))+ i s_{xy}^2\sin(\theta_{(1,2)}({\bf k})) n^+_{{\bf k}}\right)]|0\rangle    \nonumber \\
a^+_{{\bf k}}  =  \frac{d_k^+}{\sqrt{2}} \: +
\left(\: \frac{s_x  p_{kx}^+ \exp(-i\theta_x)+ s_y
p_{ky}^+\exp(-i\theta_y}{\sqrt{2} s_{xy}}\right), \nonumber  \\
n^+_{{\bf k}} =
\left(s_y \: p_{kx}^+\exp(-i\theta_x) - s_x \: p_{ky}^+ \exp(-i\theta_y)\right) / s_{xy}.
\end{eqnarray}
$\epsilon^0_{{\bf k}}=2\bar{t}_{pd}s_{xy}({\bf k})$ and $N_{k}$ are normalization factors. $\theta_1({\bf k})$ are  variationally determined:
\begin{eqnarray}
\theta_1({\bf k}) = \pm<R_I>/2\bar{t}_{pd}, ~for~state~\Theta_I, \\ \nonumber
\theta_2({\bf k}) = \pm(<R_{II}>/2\bar{t}_{pd})(\cot({\bf k}_xa)\pm\cot({\bf k}_ya)), ~for~state~\Theta_{II}.
\end{eqnarray}
These expressions are obtained from (18) and (19) for $<R>/{\bar t}_{pd} <<1$ and $k_x,k_y$ not close to zero for $\Theta_{II}$.

\subsection{Symmetries of the Mean-field TRV Hamiltonian}

It is useful to recall that a complex hamiltonian
which cannot be transformed to a real operator by any unitary (gauge)
transformation does not commute with the time-reversal operator
${\cal{R}}$ and therefore has eigenstates which break time-reversal
invariance. Correspondingly, its eigenstates cannot be transformed
real by any unitary transformation. This is true of the
wave-functions of Eq. (25,26). So Time-reversal is violated in the
phases found. Further, if
\begin{eqnarray}
H(\theta)|{k},\theta,\sigma\rangle =
\epsilon(\bf{k},\theta)|\bf{k},\theta,\sigma\rangle
\end{eqnarray}
${\cal{R}}|\bf{k},\theta,\sigma\rangle$ is not an eigenstate of
$H(\theta)$ but is of ${\cal{R}}^{-1} H(\theta){\cal{R}}$ with the same
eigenvalue $\epsilon({\bf k},\theta)$.

Consider the state $\Theta_{I}$. Physically, $\theta_1/2$ and $-\theta_1/2$ are the phase difference between the
 d-orbital and the $p_x$ and $p_y$ orbitals respectively in each unit cell. The total
  phase difference $\theta_1$ in going around any of the four triangular plaquettes
  in each unit cell is gauge invariant and can be distributed between the three legs of the
  plaquettes in any way one chooses by an appropriate gauge transformation. If we
   include additional interactions which produce additional phase shifts between the vertices of
   the plaquettes, (for example a repulsion between electrons on the neighboring oxygen orbitals
   produces in a mean-field approximation a phase shift between the $p_x$ and
   the $p_y$ orbitals of each unit cell) it can be absorbed in the invariant flux. A finite flux
   in a plaquette is equivalent to a current circulating around the plaquette.
   The current pattern for (one of the two domains of ) the state $\Theta_I$ is shown in fig.(2). This current pattern
   can be found by transforming the wave-functions to real space and calculating the expectation value
   of the current operator between the sites of the lattice. The magnitude of the current is
   of $O (\theta_1 t_{pp}t_{pd}^2a^3)$.

   The symmetry of the
   current pattern follows from the properties upon applying  the reflection operator
    $\sigma_{\hat{m}}$
   about a mirror plane $\hat{m}$ on the wavefunctions (25,26). When $\hat{m}$ is the $x=0$
   or the $y=0$ plane, the wavefunctions are eigenstates of $\sigma_{\hat{m}}$. But reflection
   symmetry is broken about the $\it{crystalline}$ mirror-planes $x=\pm y$. Note that
   if we consider properties which depend on the modulus of the wavefunctions, for example the
   charge density, mirror plane symmetries existing for $\theta =0$ are preserved for
   $\theta \ne 0$. A current sensitive experiment, (sensitive linearly to the wavefunction and its gradient) can therefore detect a Time-reversal
    breaking phase through the difference in reflection symmetries but
    experiments sensitive only to  charge-distributions cannot.

    It may also be
    seen from the wave-functions that four-fold rotation $(C_4)$ about an
    axis through the Cu's is not a symmetry but the rotation
    followed by time-reversal $C_4 {\cal R}$ is a symmetry. Also the
    center of inversion is preserved. The
    simplest {\it lattice} symmetry for the Cu-O compounds is $
    4/mmm$; the lowered symmetry generated below the transition is   $4/\underbar{m}mm$ in the double-group notation \cite{birss}.

 \begin{figure}[htbp]
\begin{center}
\includegraphics[width=0.8\textwidth]{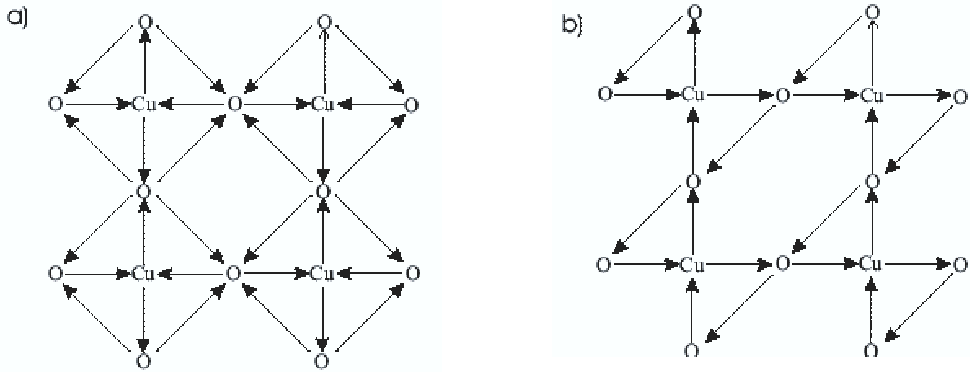}
\caption {(a): The current pattern in the time reversal violating state $\Theta_{I}$ and (b); the current pattern in the time reversal violating state $\Theta_{II}$.}
\label{fig. 2}
\end{center}
\end{figure}

In the state $\Theta_{II}$, reflection symmetry is lost about
   the $\it{crystalline}$ mirror-planes $x=0$ and $y=0$.
    The states with plus signs between the $p_x$ and the $p_y$ parts
    in the expression for $\theta_2$ keep
   reflection symmetry about $x=y$ mirror planes but not about the $x=-y$ mirror planes;
   the converse is true for the state with the minus sign. The current pattern for the
   former is also shown in fig. ( 2). Inversion is lost in this phase
   but the product of inversion and time-reversal is preserved.
   Such phases are called {\it magnetoelectric}\cite{LL}. Starting from
   {\it lattice} of symmetry $4/mmm$, the reduced new symmetry is
   $m\underbar{m}m$. These issues are further discussed in
   Ref.(\cite{dimatteo}), where extensions to more complicated
   lattice structures is also given. Note that the $\Theta_{II}$ phase has four domains.

   One may check \cite{birss} that neither the $\Theta_I$, i.e. $4/\underbar{m}mm$  nor the $\Theta_{II}$ phase, i.e.
   the $m\underbar{m}m$ is {\it piezomagnetic}; the application of a uniform magnetic field does not  linearly change the symmetry of the unit-cell.

   The polarized ARPES experiments \cite{kaminski} are consistent with the
   state $\Theta_{II}$ in region II of the phase diagram in fig. (1)\cite{borisenko}. It follows that if the superconducting transition is a continuous transition, the symmetry breaking of the pseudogap state, both time-reversal and inversion, should continue in the superconducting region \cite{agterberg,ng} to the left of the QCP in fig. (1). An experimental evidence for this would constitute a further direct test of the theory of the Cuprates presented here.

   The interesting property that chiral spin correlations accompany orbital currents of fermions ought to be mentioned \cite{wwz}. For three sites $(i,j,k)$, and  current operators $J_{ij}$ from site $i$ to site $j$, etc., the expectation value
   \begin{equation}
   \langle J_{ij}J_{jk}J_{ki}\rangle \propto \langle {\bf s}_i\cdot({\bf s}_j\times{\bf s}_k)\rangle,
   \end{equation}
   where ${\bf s}_i$ is the spin-operator at the site $i$. (The order parameter $\theta$ is proportional to this current expectation value with $(i,j,k)$ forming the nearest neighbor $O-Cu-O$ triangular plaquettes.) Therefore chiral spin order always accompanies the orbital current order.

    \section{The Fermi-surface instability of the TRV states}

   The TRV instability derived above is due to the mixing of states from different bands  (equivalently due to mixing of different correlated orbitals in a unit-cell if we start from the strong-coupling end) due to large enough screening interactions. It is not a Fermi-surface instability and by itself does not create a gap at the chemical potential.  I show here that such a TRV state cannot exist with a normal Fermi-surface. The stable state has an anisotropic gap at the chemical potential. TRV as well as translational symmetry  is preserved in the new state.

   Consider the {\it classical} fluctuation regime in the transition to the TRV breaking phase. (This should be distinguished from the regime I of the phase diagram in fig. (1))   The simplest representation of the current fluctuations in this regime has a propagator of the form:
   \begin{eqnarray}
   \chi_c({\bf q},\omega) \propto (i\omega/\gamma_0 + \kappa^2 |q|^2 + \epsilon)^{-1}
   \end{eqnarray}
   $\epsilon$ is the distance to the transition $|T_g-T|/T_g$ but the region close to the QCP at $T_g=0$ is excluded from the considerations here. Since the fluctuations are of an unconserved quantity, the leading term in their damping is a constant denoted here by $\gamma_0$. The dynamics belongs to Model A in the classification of Halperin and Hohenberg \cite{HH}.
Since the order parameter is also a discrete variable,  the fluctuations have a finite coupling to the fermions in the long wavelength limit. This is shown explicitly below. Due to the infinite range of the fluctuations, the normal Fermi-surface cannot be stable.  This is seen in a calculation of the Fermion self-energy due to exchange of the fluctuations  of Eq. (29). One finds  that for $|{\bf k-k}_f| \rightarrow 0$,
\begin{equation}
\Sigma({\bf k},\nu) = |\gamma(\hat{k}_f)|^2 \sum_{\bf q} \int_{-\infty}^{\infty}d\omega \frac{Im\chi_c({\bf q},\omega)}{\nu+\omega-\epsilon_{\bf k+q}+i\delta}(\coth(\omega/2T)+\tanh(\epsilon_{\bf k+q}/2T)).
\end{equation}
$g(\hat{k}_f)$ is the coupling function of the fermions to the long wavelength fluctuations, derived below. For a classical transition, we may take $\omega \ll T$ and consider the pole in $\chi_c$ to be much more important than the denominator in Eq. (30). Then on removing  $\omega$ from the denominator and noting that the integral of $Im \chi_c$ over $\omega$ is 0, one is left to evaluate only
\begin{equation}
\Sigma({\bf k},\nu) \approx 2T|\gamma(\hat{k}_f)|)^2 \sum_{\bf q} \frac{1}{\nu-\epsilon_{{\bf k+q}} +i\delta}
 \int_{-T}^Td\omega \frac{1}{\omega^2+(a^2q^2 +\epsilon)^2}.
 \end{equation}
 This gives for small $\nu$ and small ${\bf k-k_f}$ that
 \begin{eqnarray}
 \Sigma({\bf k},\nu) & \propto & |\gamma(\hat{k}_f)|)^2 (\nu +|{\bf k-k_f}|+\epsilon)^{-1} \ln(\nu +|{\bf k-k_f}|+\epsilon) ~~in~2~dimensions, \\
 & \propto & |\gamma(\hat{k}_f)|)^2  \ln(\nu +|{\bf k-k_f}|+\epsilon) ~~in~3~dimensions.
 \end{eqnarray}
Recall that  the self-energy due to (unscreened) Coulomb interactions in the Hartree-Fock approximation has a  $\ln|{\bf k-k_f}|$ divergence in 3d and a  $|{\bf k-k_f}|^{-1}$ divergence in 2d. Here because of the retarded nature of the effective "Coulomb" interaction, the singularities in $|{\bf k-k_f}|$ and $\nu$ are identical. The Hartree-Fock singularities  are eliminated when screened interactions are considered. In the present problem, such singularities are protected in the vicinity of the TRV transition.

The renormalized fermi-velocity is given by
\begin{equation}
{\bf v}({\bf k}_f )/{\bf v}_{f0} = z(1+{\bf v}_{f0}^{-1}\cdot \partial\Sigma/\partial {\bf q})|_{{\bf k}_f,\mu},
\end{equation}
where ${\bf v}_{f0}$ is the bare fermi-velocity and $z$ is the quasi-particle renormalization,
$z=(1+\partial \Sigma/\partial \omega)^{-1}|_{{\bf k}_f,\mu}$. Therefore the fermi-velocity is not renormalized due to the cancellation in the divergence in the frequency and momentum dependence of the self-energy. However the quasiparticle density of states at the fermi-surface is given by
\begin{equation}
N(\hat{k}_f)/N_0 = (1+{\bf v}_{f0}^{-1}\cdot \partial\Sigma/\partial {\bf q})^{-1}|_{{\bf k}_f,\mu}.
\end{equation}
Therefore the partial density of states at the fermi-surface in the irreducible representation (IR) of $|\gamma(\hat{k}_f)|)^2$ tends to 0 as $\epsilon \rightarrow 0$. If the partial density of states in any IR approaches 0, a compressibility in that IR approaches 0 signifying an instability. The cure to this instability is shown through an explicit calculation below to be a state with a gap in  the density of states in the IR in which $g(\hat{k}_f)$ is maximum.
 The connection of this situation to the Landau-Pomeranchuk instabilities is discussed in Ref.(\onlinecite{cmvpomer}).  Since the instability necessarily accompanies the classical fluctuation regime of the TRV transition, the two must occur together, as is also shown below.

  \subsection{Coupling of the Fluctuations to the Fermions}

   In this section  the coupling functions $\gamma({\bf k})$ are derived in the fluctuation regime to the TRV state and the TRV state itself. To this end, the mean-field hamiltonian, Eq. (15,16)  is supplemented by the terms giving the fluctuations in Eq. (12).

   In the $\Theta_I$  phase, the
     fluctuations  either change the magnitude of the order parameter: $(\theta_1
     \rightarrow \theta_1 + \delta\theta_{{\bf q}})$ or admix the states of the
     {\it other} domain $(\phi \rightarrow \pm \pi/2 +\delta \phi_{{\bf q}})$.
The last two terms of Eq. (12) have been used in $H_{mf}$. The first term is used to
   generate the fluctuations. To this end, one writes
 \begin{equation}
    \sum_{{\bf k,q}} A_{\bf k,q} = (\theta_1 +\delta
    \Phi_{{\bf q} }/V)
    \exp[i(\phi_0+ \delta \phi_{{\bf {q}} })] ~
   + ~ Incoherent~part.
    \end{equation}
    The incoherent part is the fermionic part of the fluctuations
    which is expressed in terms of
    the new particle-hole states in the $\Theta$ phases. The coherent part
    are the bosonic fluctuations. In the fluctuation regime to the $\Theta$ states, $\theta=0$ and the only fluctuations are the amplitude fluctuations $\delta \theta_{{\bf q}}$ in the simple small fluctuation approximation used here.

The Fermion-Boson coupling hamiltonian follows directly by finding the matrix elements of the fermionic part of $A_{\bf k,q}$ in
 the wavefunctions given in Eqs. (25,26).  The results are
  \begin{equation}
 H_{FB} = \sum_{{\bf k},{\bf k}'} c_{{\bf {k}}',\theta}^{\dagger}
 c_{{\bf k},\theta}\left(\gamma_p({\bf{k}},{\bf{k}}')[\delta\theta_{({\bf k}-{\bf k}')}
   +\delta\phi^{\dagger}_{-({\bf {k}}-{\bf{k}}')}] +
   \gamma_a({\bf{k}},{\bf{k}}')[\delta\theta_{({\bf k}-{\bf k}')}
   +\delta\theta^{\dagger}_{-({\bf {k}}-{\bf{k}}')}]\right)
  \end{equation}
    In the next section we will be interested in the coupling only in the
   forward scattering limit $({\bf{k}}-{\bf{k}}'\rightarrow 0)$. In this limit, for the $\Theta_1$ phase,
  \begin{equation}
  \gamma_{a}({\bf{k}},{\bf{k}}) \approx V\theta_1 \left(1+\cos( k_xa)+
  \cos( k_ya)\right),  \nonumber  \\
  \gamma_{p}({\bf{k}},{\bf{k}}) \approx V\theta_1 \left(\cos( k_xa)-
  \cos( k_ya)\right).
   \end{equation}

The energy of the phase mode at long wavelengths is calculated in
Appendix A to be
\begin{eqnarray}
\Omega_{\phi} \approx\frac{\bar{t}_{pp}^2}{\bar{t}_{pd}}\theta_1^2.
\end{eqnarray}
The energy of the amplitude mode can be shown to be higher by a numerical factor of
$0(2)$.

In the fluctuation regime, the only coupling is  to the amplitude mode.  The leading forward scattering coupling  has the dependence,
\begin{equation}
\gamma_{fluct}({\bf{k}},{\bf{k}}) \approx V/2 \left(\cos( k_xa)-
  \cos( k_ya)\right).
  \end{equation}

   In the $\Theta_{II}$ phase the fluctuations may change the magnitude of the
   order parameter $(\theta_2
     \rightarrow \theta_2 + \delta\theta_{{\bf q}}$, admix the states of the
     {\it other} time-reversed phase $\phi \rightarrow \pm \pi/2 +\delta \phi_{{\bf q}}$ ,or
     admix the states of the other {\it spatial} domain, i.e. admix states  with reflection symmetry
     about the $(\hat{x}+\hat{y})$  if the
     mean-field state chosen is the one with reflection symmetry about $(\hat{x}-\hat{y})$
     or vice-versa. We will call the operator
     for the last $\delta \theta_{m,{\bf q}}$. So in addition to
     the two kinds of coupling in Eq. (37), a coupling to
     fluctuations $\delta \theta_{m,{\bf q}} + \delta \theta^{\dag}_{m,{\bf
     q}}$ with a coupling function $\gamma_m({\bf k+q},{\bf k})$ must be included.

     To ensure a gauge-invariant calculation in the $\Theta_{II}$ phase, a hamiltonian
     with a larger symmetry than that used to generate the mean-field
     state must be used. A minimum such Hamiltonian includes both the
      $A_3^{\dagger}A_3$ and $A_4^{\dagger}A_4$ terms of Eq. (9). We can then follow
      the familiar procedure to generate the coupling functions in
      the forward scattering limit for the three modes already described:
      For the amplitude mode, phase mode and the "mixing mode" we get, respectively
   \begin{eqnarray}
   \gamma_a({\bf k},{\bf k}) \approx V \theta_2 (\cos(k_xa) + \cos(k_ya))  \\
   \gamma_p({\bf k},{\bf k}) \approx V \theta_2 (\sin(k_xa) \pm \sin(k_ya))  \\
   \gamma_m({\bf k},{\bf k}) \approx V \theta_2  (\cos(k_xa) - \cos(k_ya)).
   \end{eqnarray}

   The energy of these modes in the long wavelength limit are estimated in the
   Appendix. The energy of the phase mode $\Omega_{\phi}$ is about
   the same as that in the phase $\Theta_I$. the energy of the
   mixing mode at long wavelengths $\Omega_m$ is estimated in the appendix to be about 1/2 of these.
  Therefore the important
   mode to consider for stability in the state $\Theta_{II}$ is  the mixing mode.

   The forward scattering with the amplitude fluctuations in the fluctuation regime has a dependence,
   \begin{equation}
   \gamma_{fluct}({\bf k},{\bf k}) \approx V/2 (\sin(k_xa) \pm \sin(k_ya))
   \end{equation}

   I have only considered small fluctuations in the above. The model also has interesting topological fluctuations (mentioned in the concluding section.)

 \subsection{Anisotropic Gap (Pseudogap)  in the TRV state}

 In the first part of this section, I presented an argument from $T\gtrsim T_g$ that the TRV phase has a tendency to develop an anisotropic gap at $T=T_g$. Now, I present a calculation coming to the same conclusion from $T\lesssim T_g$. In effect, the zero order  collective modes in the TRV phase, whose frequencies are estimated above, are shown to be unstable due to renormalizations by $H_{FB}$ with a normal fermi-surface.

 The effective Hamiltonian  in  the TRV phase
 is
 \begin{equation}
 H_{eff}^{trv} = \sum_{\bf{k}\sigma}\epsilon({\bf {k}},\sigma)
 c^{\dagger}_{{\bf {k}}, \sigma}c_{\bf {k},\sigma}
 +\sum_{\bf{q}}\Omega({\bf{q}})b^{+}({\bf{q}})
 b({\bf{q}}) +H_{FB}.
 \end{equation}
 $H_{FB}$ is given by Eq.(37). The annihilation, creation operators of the  fluctuation modes are denoted  by $b,b^+$.
 The fluctuations with the lowest energies are of
 interest. They are the fluctuations of the phase for the
 $\Theta_I$ phase and the "mixing"-fluctuations for the
 $\Theta_{II}$ phase. The operators $b({\bf q}),b^+({\bf q})$
 will only refer to these.
 We may eliminate these
  fluctuations and $H_{FB}$ to generate a retarded effective interaction between the Fermions
  of the form
  \begin{eqnarray}
  \sum_{{\bf k,k}^{\prime},{\bf q},\sigma,\sigma^{\prime}}
\gamma({\bf k,k+q})\gamma({\bf k}^{\prime}-{\bf q,k}^{\prime})\chi_c(q,\omega)
c_{{\bf k+q}  \sigma}^+c_{{\bf k } \sigma}
c_{{\bf k}^{\prime}-{\bf q} \sigma^{\prime}}^+
c_{{\bf k}^{\prime} \sigma^{\prime}}.
\end{eqnarray}
where $\chi_c(q,\omega)$ is the propagator for the relevant fluctuations.
$\chi_c$ may be
approximated by $-2/\Omega_q^0$
with a cut-off such that
$\epsilon(k^{\prime}),\epsilon(k^{\prime}+q)$
are both within about $\omega_c$ of the chemical potential. The effective interaction  is repulsive for states
 at higher energies. We use only the forward scattering part of
 the interactions, i.e.
 in the limit of small ${\bf{q}}$,and for ${\bf{k},{k}'}$ close to the Fermi-surface:
 \begin{eqnarray}
  g({\bf {k}, {k}'})= - Lim_{{\bf q}\rightarrow 0} 2 \gamma({\bf k,k+q})\gamma({\bf k}^{\prime}-{\bf q,k}^{\prime})/\Omega_{{\bf{q}}},
  \end{eqnarray}
  with a cut-off of order $\omega_c \approx \Omega_0$. If $\Omega_0 \to 0$, the cut-off $\omega_c$ is approximately the larger of $\Omega_0$ or the damping of the collective mode $\gamma_0|\epsilon$ of Eq.(29).

 The instability seen through Eq. (30) may  also be seen through the equation of motion for the { \it fluctuations} $<c_{{\bf{k}}+{\bf{q}}, \sigma}^+c_{{\bf{k}}, \sigma}>$. Just as for Landau-Pomeranchuk instabilities,
  no solution for such equations can be found for real frequencies in the limit of small ${\bf q}$ \cite{pines-nozieres, cmvpomer}.
  The cure to this instability  is  a new state
  in which particles are annihilated
 and created by operators $\tilde{c}^{\dagger}_{{\bf{k}},\sigma},\tilde{c}_{{\bf{k}},\sigma}$
 respectively,
 such that a stable homogeneous equation for
 \begin{eqnarray}  \langle
  \tilde{c}^{\dagger}_{{\bf{k}+{q}},\sigma'}\tilde{c}_{{\bf{k}},\sigma}\rangle({\bf{q}},\nu) .
  \end{eqnarray}
at $\nu=0$ can be found at ${\bf q}$.  Correspondingly, in the stable state,
the one-particle eigenvalues are  renormalised from
$\epsilon_{{\bf{k}} \theta \sigma}$ to $E_{{\bf{k}}\sigma}$.
Actually, the interactions $\gamma({\bf k,k'})$ are renormalized
also. But such strong-coupling corrections are not considered here. The particular form for
$E_{{\bf{k}}\sigma}$ and the relation of $\tilde{c}$'s to $c$'s,
etc. is dictated by the self-consistency condition
derived below.

 A stable state is chosen in which  $  <c_{{\bf{k}+{q },\sigma}}^+c_{{\bf k },\sigma}>$, and  $ <b_{{\bf q}}>, <b^+_{{\bf q}}>$ acquire non-zero values for small ${\bf q}$. ( In taking the limit, ${\bf k+q}$ and ${\bf k}$ should be on opposite sides of the Fermi-surface; the limit ${\bf q} \rightarrow 0$ of $<c_{{\bf{k}+{q },\sigma}}^+c_{{\bf k },\sigma}>$ is not the particle density at ${\bf k}$.)  The self-consistency equations are derived in a mean-field approximation as follows:
By a mean-field decomposition and minimization of  the effective hamiltonian, Eq. (44),
 \begin{equation}
 \langle b_{{\bf q}}\rangle = -\sum'_{{\bf k}\sigma} \frac{\gamma({\bf k},{\bf
 k+q})}{2\Omega_{{\bf q}}}\langle c^+_{{\bf
 k+q}\sigma}c_{{\bf k}\sigma}\rangle
 \end{equation}
 The restriction on the sum indicates that the sum is over states within $\omega_c$ of the chemical potential and that the states ${\bf k,k+q}$ refer to (occupied, unoccupied) states or vice-versa.

 The effective {\it mean-field} hamiltonian for the electrons is
 now
 \begin{eqnarray}
 H_{eff} & = &\sum_{{\bf k}\sigma}\epsilon_{{\bf k}}c^+_{{\bf k}\sigma}c_{{\bf
 k}\sigma} + \sum'_{{\bf k,q}\sigma}V({\bf k,k+q})c^+_{{\bf
 k+q}\sigma}c_{{\bf k}\sigma} \equiv \sum_{{\bf k}\sigma} E_{{\bf k}}\tilde{c}^+_{{\bf k}\sigma}\tilde{c}_{{\bf k}\sigma} \\
V({\bf k,k+q}) & \equiv & \gamma({\bf k},{\bf k+q})(\langle b_{{\bf
q}}\rangle +\langle b^+_{-{\bf q}}\rangle).
\end{eqnarray}

One may write $H_{eff} $ in real space as a tight binding Hamiltonian on the lattice,
\begin{equation}
    H_{eff} = \sum_{{\bf R}_i,{\bf R}_j\sigma} t({\bf R}_i,{\bf R}_j) c^+_{i\sigma} c_{j\sigma},
\end{equation}
which shows {\it effective} long-range, angle-dependent hopping, which is to be determined self-consistently.

The one-particle Green's function for $H_{eff}$ in the
self-consistent Brillouin-Wigner approximation  has a self-energy which is an infinite continued fraction which may be summed formally to give the one-particle propagator
\begin{eqnarray}
G({\bf k},\omega) = \frac{1}{\omega -\epsilon_{{\bf k}}- D({\bf
k},\omega)},
\end{eqnarray}
where
\begin{eqnarray}
D({\bf k},\omega) = \sum'_{{\bf q}} \frac {|V({\bf k,k+q})|^2}
{\omega -\epsilon_{{\bf k+q}}- D({\bf
k+q},\omega)}.
\end{eqnarray}
 On the energy shell, at
 \begin{equation}
 \omega = E_{{\bf k}}= \epsilon_{{\bf k}}+ D({\bf
k},E_{{\bf k}}),
\end{equation}
\begin{eqnarray}
D({\bf k},E_{{\bf k}}) = \sum'_{{\bf q}} \frac {(f(E_{{\bf
k+q}})-f(E_{{\bf k}}))|V({\bf k,k+q})|^2}{E_{{\bf k+q}} - E_{{\bf
k}}} .
\end{eqnarray}
This formal solution to (50) is exact in the limit that the number of states $({\bf k+q})$ coupled to a given state ${\bf k}$ is very large compared to 1. It also follows by taking the expectation value of (50) in the new state that
\begin{eqnarray}
D({\bf k},E_{{\bf k}})= \sum_{{\bf q}}V({\bf k,k+q})\langle c^+_{{\bf
 k+q}}c_{{\bf k}}\rangle.
\end{eqnarray}
Comparing Eqs. (56) and (57) and using Eq.(51),
\begin{eqnarray}
\langle c^+_{{\bf
 k+q}}c_{{\bf k}}\rangle = \frac{\gamma({\bf k},{\bf k+q})(f(E_{{\bf k+q}})-f(E_{{\bf
 k}}))(\langle b_{{\bf q}}\rangle+\langle b^+_{-{\bf q}}\rangle}{E_{{\bf k+q}} - E_{{\bf k}}}.
 \end{eqnarray}
  Inserting $\langle b_{{\bf q}}\rangle,\langle b^+_{-{\bf q}}\rangle$ from Eq. (49) and after a simple manipulation
  I obtain the self-consistency condition  :
\begin{eqnarray}
V({\bf{k+q/2},{k-q/2}}) = \sum '_{\bf{k}'}
g({\bf k,k}')\frac{f(E_{{\bf{k}}'+{\bf{q}}/2})
-f(E_{{\bf{k}}'-{\bf{q}}/2})}{(E_{{\bf{k}}'+{\bf{q}}/2}-E_{{\bf{k}}'-{\bf{q}}/2})}
 V({\bf{k'+q/2}},{\bf{k'-q/2}}),
\end{eqnarray}
 In Eqs. (59), the sum is restricted to states ${\bf k}'$
 such that $\epsilon({\bf k}')$ is within about $\Omega_0$ of the chemical potential $\mu$,
  as required by the retarded nature of the Fermion-Boson interaction.

  Note that we have already proven in Eq. (55)  that the quasi-particle energies in  the new state obey,
\begin{eqnarray}
E_{{\bf k}\sigma} = \epsilon_{{\bf k}\sigma}+D_{{\bf k}\sigma}.
\end{eqnarray}
There is no self-energy in quadrature as in the BCS theory or the theory
of CDW/SDW. This of-course has pronounced effects on the spectroscopic and thermodynamic properties in the new state.

\subsection{Solution near $T_g$}

The self-consistent solution of Eqs. (56),(58) and (59) is a complicated problem which however simplifies in the "Ginzburg-Landau" regime which provides the transition temperature and the leading temperature dependence of the gap below it. In this regime only very small ${\bf q}$ old states are admixed to a given ${\bf k}$ in the new states. I will assume that the solution obtained continues qualitatively to lower temperatures.

In this regime, note first that since $g({\bf{k},{k}'})$ is
 proportional to $\gamma({\bf{k}})\gamma({\bf{k}'})$
for $q \rightarrow 0$, $V(k,k+q) \sim \gamma({\bf{k}})$ also.
  Second, $E(k) \approx \epsilon(k)$ for $|\epsilon(k) - \mu | \gtrsim \Omega_0^0$.
The self-consistency condition (59) may therefore be rewritten in the limit ${\bf{q}}
 \rightarrow 0$ as
 \begin{eqnarray}
 1 =   \frac{-2}{\Omega_0}\sum ^{'}_{{\bf{k}}}|\gamma({\bf{k}})|^2\left(
  \frac{f(E^{>}_{{\bf{k}}})-
  f(E^{<}_{{\bf{k}}})}{E^{>}_{{\bf{k}}}
   -E^{<}_{{\bf{k}}}}\right).
  \end{eqnarray}
Here $E^{>}_{{\bf{k}}},E^{<}_{{\bf{k}}}$ are the new one particle
eigenvalues above and below the chemical potential respectively.  To determine $E_{{\bf k}}$, one must satisfy Eq. (60) in a manner continuous with the approach to the instability, where $\partial \Sigma/\partial {\bf k}$ diverges. Accordingly, I seek a solution with
 \begin{eqnarray}
  E^{>}_{{\bf{k}}} = \epsilon_{\bf{k}} + D(\hat{k}_f), \nonumber \\
  E^{<}_{{\bf{k}}} = \epsilon_{\bf{k}} - D(\hat{k}_f), \nonumber \\
  (E^{>,<}_{{\bf{k}}} - \epsilon_{\bf{k}}) \rightarrow 0~for ~| E^{>,<}_{{\bf{k}}}-\mu| >> \Omega_0.
 \end{eqnarray}
 with the additional condition that $D(\hat{k}_f)\geq 0$ for all $\hat{k}_f$.

 This manner of opening a gap at the
 chemical potential
 appears is new. The physics of this gap is distinct from that of the
   gaps of Bloch-Wilson one-electron theory,
 the BCS gap,
 the Mott gap or the charge-density (CDW)or spin-density wave (SDW) gaps. In the mean-field approximation the gap arises due to an effective infinite range hopping exhibited in Eq. (52).

The integral equation (61) for $D({\bf k})$ may be rewritten as
\begin{eqnarray}
 1 = -\frac{-2}{\Omega_0}\int \frac{d{\hat{k_f}}}{2\pi}\frac{|\gamma({\hat{k}_f})|^2}{
 D(\hat{k}_f)} \int_0^{\omega_c}
  d\epsilon \rho_0(\epsilon) \frac{tanh(\beta D(\hat{k}^f))} {1+cosh(\beta \epsilon)sech(\beta D(\hat{k}^f)}.
  \end{eqnarray}

At temperatures for which $\beta D(\hat{k}_f) \ll
1$, Eq. (63) reduces to
\begin{eqnarray}
1 -tanh(\beta \omega_c/2) \int \frac{1}{\Omega_0}\rho(0) \frac{d{\hat{k_f}}}{2\pi}
 |\gamma(\hat{k})|^2 = \\ \nonumber
 - \beta ^2 /12 \left(tanh(\beta \omega_c/2) -
 tanh^3(\beta \omega_c/2)\right) \frac{1}{\Omega_0}\rho(0)\int \frac{d\hat{k}_f}{2\pi}
 |\gamma(\hat{k})|^2 D^2(\hat{k}_f).
\end{eqnarray}
This constitutes the microscopic derivation of a free-energy of the Landau form for the instability.

 It is interesting to make connection with the forward scattering parameters of Landau theory. This may be done only when the cut-off $\omega_c$ satisfies $\beta \Omega_0 >>1$. In that case define,
 \begin{eqnarray}
-\frac{1}{\Omega_0}\rho(0)\int
\frac{d{\hat{k_f}}}{2\pi}|\gamma(\hat{k}_f)|^2  \equiv F_{\ell}^s/(2\ell+1),
\end{eqnarray}
 where $F_{\ell}^s$ is the spin-symmetric Landau parameter in the $\ell$-th "angular momentum" channel. The condition for the instability is that the left side of Eq. (64)
 be zero. This happens  at a temperature $T_{\ell}$ below which
 stability is achieved with a finite value for the right side, i.e. with $D(\hat{k},T)\ne 0.$
 The transition temperature in that case is
  \begin{eqnarray}
  T_{\ell} \approx \frac{\omega_c}{\ln\left(\frac{F_{\ell}^ s/(2\ell+1)}{1+F_{\ell}^
  s/(2\ell+1)}\right)}.
\end{eqnarray}

In our problem we must take the limit, $\beta \omega_c \ll 1$ for $T$ just below $T_g$, where $\Omega_0 \to 0$. Then the cut-off $\omega_c \approx \gamma_0|\epsilon|$, defined in Eq.(29). The transition temperature $T_g'$ is then given by
\begin{eqnarray}
T_g' \approx (\rho_0/2)\gamma_0|\epsilon|\int  \frac{d{\hat{k_f}}}{2\pi}\frac{|\gamma(\hat{k}_f)|^2}{\Omega_0} \approx \left(\frac{\rho_0 V^2\gamma_0}{\bar{t}_{pp}^2/\bar{t}_{pd}}\right)\left(1-T_g'/T_g\right).
\end{eqnarray}
Here the relationship, Eq.(39) between the collective mode frequencies $\Omega_0$ and the order parameter $\theta$ has been used. Eqn.(67) gives $T_g'\approx T_g$ for $(\rho_0V^2\bar{t}_{pd}\gamma_0)/(T_g\bar{t}_{pp}^2) \gg1$. For $\gamma_0 \approx \rho_0^{-1}$ and $x \approx 0.1$, this ratio is of $O(10^4)$. One should expect that in a better theory, the TRV instability is accompanied exactly by the instability to the anisotropic gapped state. As explained later, the mean-field theory of the TRV transition is expected to be strongly modified by fluctuations. The results for the region near the transition obtained here should only be regarded as indicative.

The value of the gap for $T<T_g$ is also given by Eq.(64):
\begin{equation}
\frac{\int \frac{d{\hat{k_f}}}{2\pi}|g\gamma(\hat{k})|^2 D^2(\hat{k})}
{\int  \frac{d{\hat{k_f}}}{2\pi}|\gamma(\hat{k})|^2}\approx 6 T_g(T_g-T).
\end{equation}
 The angular dependence of $D(\hat{k})$ is to be found from the equation (68), subject to the condition that $D^2(\hat{k}) >0$ for all $\hat{k}$. A family of  solutions to (68) exist, the one that must be chosen is that which minimizes the energy by considering the fourth order terms in the free-energy. This is given by $D(\hat{k})$ which has the same dependence as  the kernel
$|g(\hat{k}_f)|^2 $,
\begin{equation}
D(\hat{k}) \propto    |\gamma(\hat{k}_f)|^2 .
\end{equation}
Referring to Eqs. (38,40), in the $\Theta_I$ phase, $ |g(\hat{k}_f)|^2 \propto
\left(\cos(k_{fx}a)-\cos(k_{fy}a)\right)^2.$ (This is also true in the fluctuation regime to the  However in the fluctuation regime to the $\Theta_{II}$ phase, using Eq. (44),
$ |g(\hat{k})|^2 \propto
\sin^2(k_{fx}a/2\pm k_{fy}a/2),$ the two signs refer to the two different reflection symmetry violating domains of $\Theta_{II}$.

In the $\Theta_{II}$ phase, the most unstable mode  is the mixing mode for which the coupling function again has "d-wave symmetry' and the gap is expected to have the shape  $(\cos(\hat{k}_{fx})-\cos(\hat{k}_{fy}))^2$.

The value of the gap  is expected to  grow for lower temperatures  approximately as in Eq.(68). The amplitude of the gap in the one particle spectrum at the
chemical potential $D_0$ at $T=0$ is $\approx \sqrt{6} T_g(x)$.

One should ask at this point  how the transition to the TRV symmetry is affected by the pseudogap appearing simultaneously with it. Recalling that the transition to that phase depends in admixing states
between the bonding and the non-bonding bands, the effect is only of order $D_0/t_{pd} <<1.$
It would of-course be desirable to derive the TRV plus the anisotropic gap in a single mean-field ansatz
rather than with the procedure used here.

\section{Some Properties of the TRV-Pseudogap State}

In (62), the dispersion $E_{{\bf k}}$ is determined for $\beta D(\hat{k})\ll 1$. To calculate properties at intermediate and low  temperatures we need $E_{{\bf k}}$ for $\beta D(\hat{k})\gtrsim 1$. To meet the requirement  that $E_{\bf k} \rightarrow \epsilon_{\bf k}$ for $|E_{\bf k}-\mu| \gg D(\hat{k})$, we may choose
\begin{eqnarray}
E_{\bf k}=\epsilon_{\bf k} + D(\hat{k}) /(1+(\epsilon_{\bf k}/\epsilon_c)^2).
\end{eqnarray}
 The derivation of $\epsilon_c$ is only possible through a solution of the self-consistency equations at low temperatures together with the requirement that the number of states below the chemical potential is constant. One expects $\epsilon_c$ to be $O(D_0)$. The dispersion in Eq. (70) is represented in fig. (3).

\begin{figure}[htbp]
\begin{center}
\includegraphics[width=0.8\textwidth]{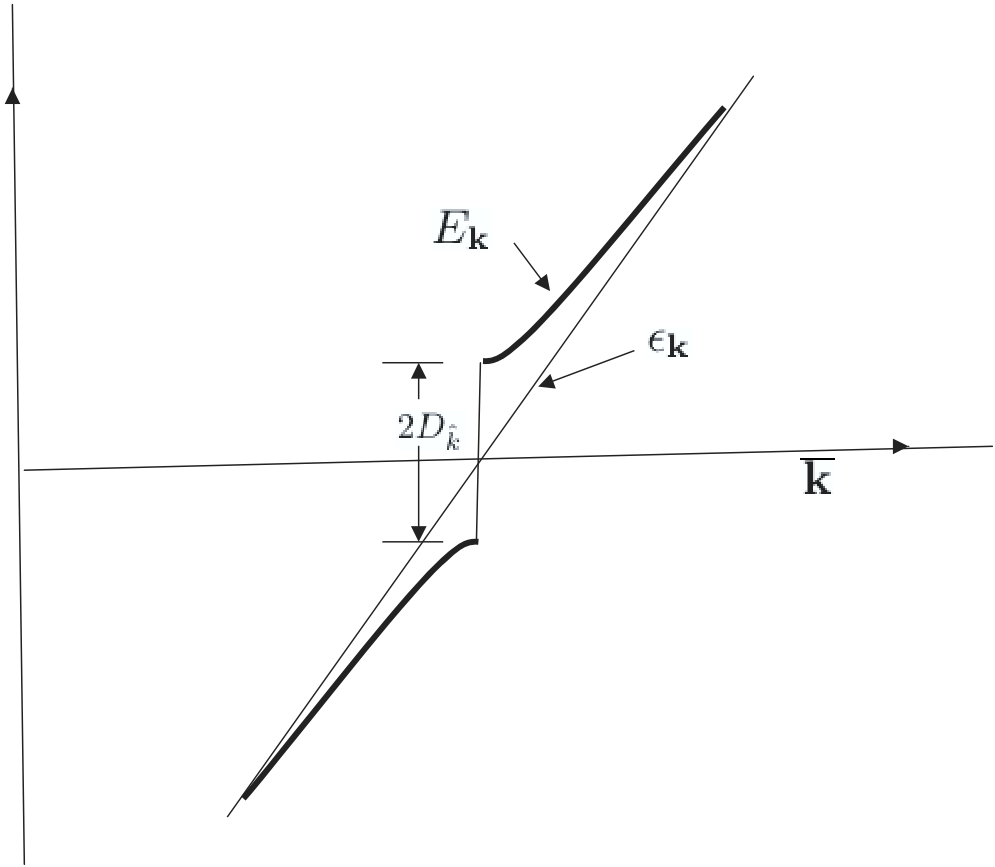}
\caption{The one-particle dispersion measurable by ARPES expected for the pseudogap phase. $\epsilon_c$ has been assumed to be $=D_0$, the maximum gap. The discontinuity D is a function of angle around the Fermi-surface. }
\label{fig. 3}
\end{center}
\end{figure}

 The density of the state in the pseduogap phase may be calculated as
\begin{eqnarray}
N_p(\omega)  = \sum_{\bf k} (1/\pi) \delta(\omega-E_{\bf k}).
\end{eqnarray}
If $\epsilon_c$ were $ \gg D_0$, this can be evaluated analytically to give
 \begin{eqnarray}
N_p ( \omega ) &=& \nu (0) \frac{2}{\pi} arcsin
\left( \left| \frac{\omega}{D} \right|^{1/2} \right) ,
\left| \frac{\omega}{D} \right| \leq 1 \nonumber \\
&=& \rho (0) , \left| \frac{\omega}{D} \right| \geq 1
\end{eqnarray}
This increases as $\left| \omega / D \right|^{1/2}$ for
$\left| \omega / D \right| << 1$.
For a general $\epsilon_c$, the asymptotic low energy density of states keeps this form.
For $\epsilon_c$ of  $O( D_0)$, the density of states can only be evaluated numerically.

This density of states ignores both the effect of impurity scattering as well as  lifetimes due to inelastic scattering which are a function of energy and temperature. The latter are included below in calculating the single-particle spectral function.

 \subsection{Single-particle Spectral Function}

 \subsubsection{Inelastic Scattering}

To obtain the spectral function $A (k, \omega )$
measured by ARPES
one needs besides Eq.(70), the
self-energy of the single particle states.
To calculate the scattering rate, we first estimate the polarizability
$\chi_0 ({\bf k,q}, \omega )$ from a single-particle-hole bubble formed from states ${\bf k}$ and ${\bf k+q}$.

\begin{eqnarray}
Im \chi_0 ({\bf k,q}, \omega ) \propto \sum_{\bf k} \delta\left(\omega- (E_{\bf k+q} -E_{\bf k})\right) [f(E_{\bf k+q})-f(E_{\bf k})]
\end{eqnarray}
We are interested especially in the low energy contributions to $Im \chi_0 ({\bf q,k} \omega )$. These come when the initial state ${\bf k}$ is near a zero of the pseudogap. With such an initial state and a the gap in the single-particle spectrum  $D(\hat{{\bf q}})$, the integrals in Eq.(73) may be evaluated to yield the result that

\begin{eqnarray}
Im \chi_0 ({\bf k,q}, \omega ) \propto \omega \Theta(\omega-D(\hat{{\bf q}}))/v_f^2
\end{eqnarray}
with a function of the angle ${\bf q}$ multiplying (74) which has a weak dependence, except when ${\bf k+q}$ is also near one of the zeroes of the gap function. In that case
\begin{eqnarray}
Im \chi_0 ({\bf k,q}, \omega ) \propto \propto \omega/q^2 \Theta(\omega-D({\hat{\bf q}})).
\end{eqnarray}

Given that the spectral weight  of fluctuations for most $({\bf k,q})$ is 0 below an energy $D(\hat{{\bf q}})$ and increases only linearly beyond it, no important modifications are expected due to interactions.
  For
$\omega \gg D (q)$, $Im \chi_0 ({\bf k,q}, \omega)$ must have the same form as in the normal state
for all ${\bf k}$, i.e. of the marginal Fermi-liquid hypothesis.

From the estimates above one can calculate  the form of the self-energy $\Sigma({\bf k},\omega)$ for ${\bf k} $ in the vicinity of the region of zero gap. For this calculation the region in which $Im\chi$ is given by (75) is ignorably small. Using (74), a simple calculation gives for  $\omega/D_0\ll1$ that

\begin{eqnarray}
Im\Sigma({\bf k},\omega) \propto  N(0)\omega^2/D_0.
\end{eqnarray}
For  $\omega/D_0\gg1$, the self-energy  must revert to the MFL form.

For a ${\bf k}$ away from the zero of the gap, and near energy $D({\hat{{\bf k}}})$, no decay through an inelastic process is possible kinematically for $\omega$ less than $D({\hat{{\bf k}}})$. (This is the process
where the intermediate Fermion state is near the zero of the pseudogap and the scattered polarizability
makes up the momentum difference.) Therefore for
$ ( \omega , T ) << D ({\hat{\bf k}})$ the self-energy  is exponentially small. For
$( \omega , T ) \gtrsim D ({\hat{\bf k}})$
it returns to the value $\Sigma_n( \omega , q , T)$ without the
pseudo-gap.  So
\begin{equation}
Im \Sigma ( \omega , {\bf k}, T) \approx \; \mbox{sech} \;
\left( \frac{D ({\hat{\bf k}},T)}{( \omega^2 + \pi^2 T^2 )^{1/2}}\right)
Im \Sigma_n ( \omega , {\bf k}, T) \;.
\end{equation}

The spectral function at the wave-vectors away from the zeroes of the gap
$E(\hat{k}_f)= \pm D(\hat{k}_f)$ calculated using Eq. (77)  and
the marginal Fermi-liquid form for $\Sigma_n$ is plotted in fig.(4) for
a few temperatures.  A 'pseudogap' in the direction $\hat{k}_f$ appears
below $T \approx D(\hat{k}_f)$.  At zero temperature the Fermi-surface is composed of four points. But at finite temperature a gap can be discerned in a direction ${\hat{\bf k}}_f $ only if $T$ is less than about $D(\hat{k}_f, T)$. This produces the illusion of `Fermi-arcs', which shrink as the temperature decreases.

  \begin{figure}[htbp]
\begin{center}
\includegraphics[width=0.8\textwidth]{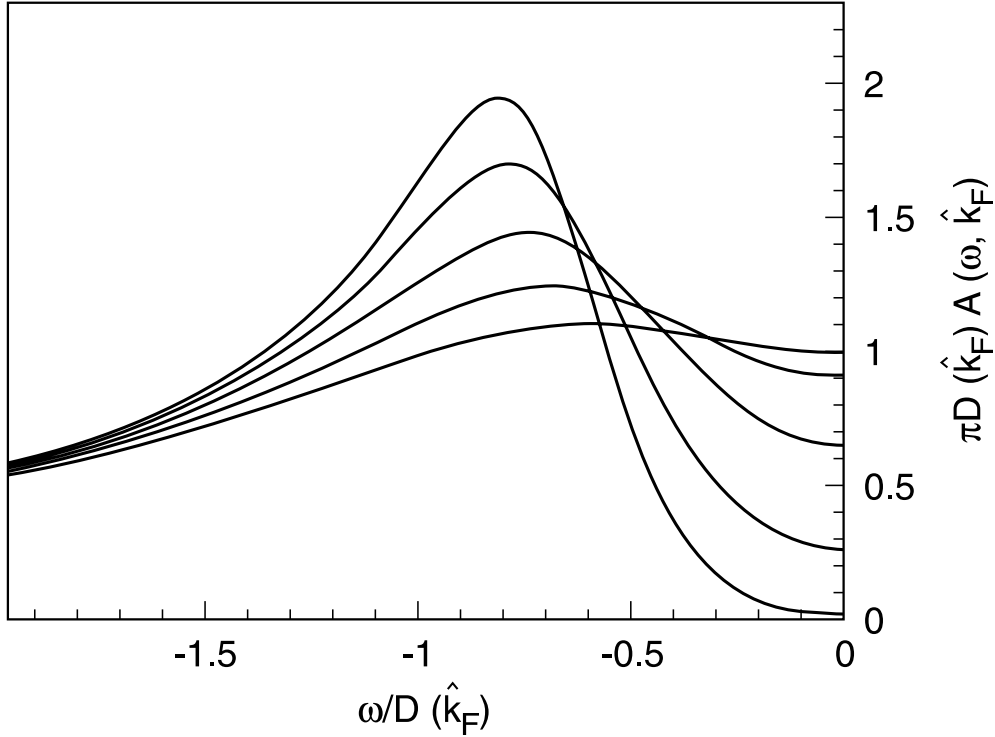}
\caption{Calculated Spectral function in the pseudogap phase excluding linewidth due to impurity scattering. The spectral function in any given direction $\hat{k}$ has been normalized to the value of the gap in that direction $D{\hat k}$ and the energy is also normalized to the value of the gap. The temperatures shown are given by $T=nD(\hat{k}_f)/4\pi$ for n=1,2,..5.}\label{fig. 4}
\end{center}
\end{figure}

\subsubsection{Impurity Scattering}

It is important in the Cuprates (and other layered materials) to distinguish the effect of impurities in the planes which are expected to acts as point or large-angle scatterers
from that of impurities in between the planes which scatter electrons in the planes only through small angles on the Fermi-surface \cite{abrahams-varma-pnas, abrahams-varma-hall}. In the latter case, the scattering rate at a point ${\bf k}$ (calculated in the Born approximation) depends on the local density of states at the point, i.e. on the spectral function $A({\bf k},\omega)$. This idea is important to understand the difference in the scattering rate deduced from single-particle spectra from that in transport, the insensitivity of $T_c$ to small angle scattering, the Hall effect and the lineshape as a function of $\omega$ and its variation with $\hat{k}_f$ of the spectral function $A({\bf k},\omega)$  in the superconducting state \cite{hirshfeld-small angle}.

 Given the form of the spectral function for the pseudogap {\it neglecting impurity scattering} in fig. (4), small angle impurity scattering also has interesting effects on the {\it observable} spectral function in the pseudogap state. It follows that elastic scattering will be anisotropic, increasing from the nodal region where the spectral function is small to the antinodal region, where it is large. It also follows that the small angle scattering rate will be small in the antinodal region in the tail of  $A({\bf k},\omega)$ towards the chemical potential and large on the other side. Such features have been observed in experiments \cite{kaminski-pr}. The detailed comparison of experimental $A({\bf k},\omega)$ with calculations should incorporate these effects of impurity scattering.

 In the classical fluctuation regime I, above the pseduogap regime, the single-particle spectra should also show the frequency dependence calculated in Eqs.(31,32). Since, it is only weakly dependent on $\nu,|k-k_f|$ and $\epsilon$, it may appear within the experimental uncertainty as enhanced elastic scattering, which is angle dependent $\propto |\gamma(\hat{k}_f)|^2$.

\section{Comparison with measured properties of the pseudogap phase}

           The effect of impurities is much less in evidence when one calculates the density of state $ N_p(E)$ by integrating the spectral function over ${\bf k}$. The calculated $N_p(E)$, for any $x$ at several  temperatures below $T_p(x)$ is shown in fig.(5).   To compare, the measured density of states by STM is shown in fig. (6). I do not have anything to say about the general approximately linear variation of the density of states as a function
  of energy over the whole energy range, which is found both above and below $T_p$. If one looks at the features brought about by the pseudogap, good correspondence may be discerned in both the energy and the temperature dependence. Of special importance is to note the absence of the BCS type singularity in the density of states in the pseudogap both in the calculations and  in the experiments in the pseudogap regime. In the experiments it appears as one cools below $T_c(x)$. (Such singularity is present in the pseudogap region in theories based on  any translational symmetry breaking, for example a charge density wave or a D-density wave.) The variation of the pseudogap feature as a function of temperature in the calculated curves also bears close correspondence to the experiments.
    \begin{figure}[htbp]
\begin{center}
\includegraphics[width=0.8\textwidth]{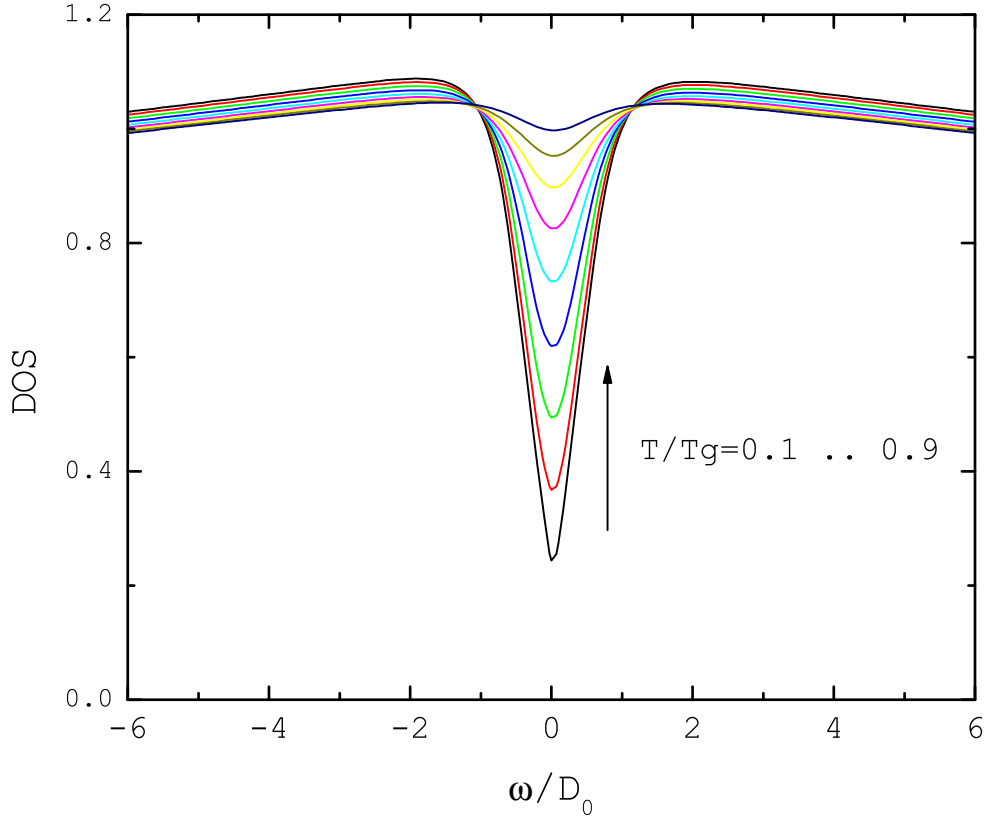}
\caption{Calculated density of states in the pseudogap phase at several temperatures. $D_0$ is chosen 2.5$ T_g$}
\label{fig. 5}
\end{center}
\end{figure}

  \begin{figure}[htbp]
\begin{center}
\includegraphics[width=0.8\textwidth]{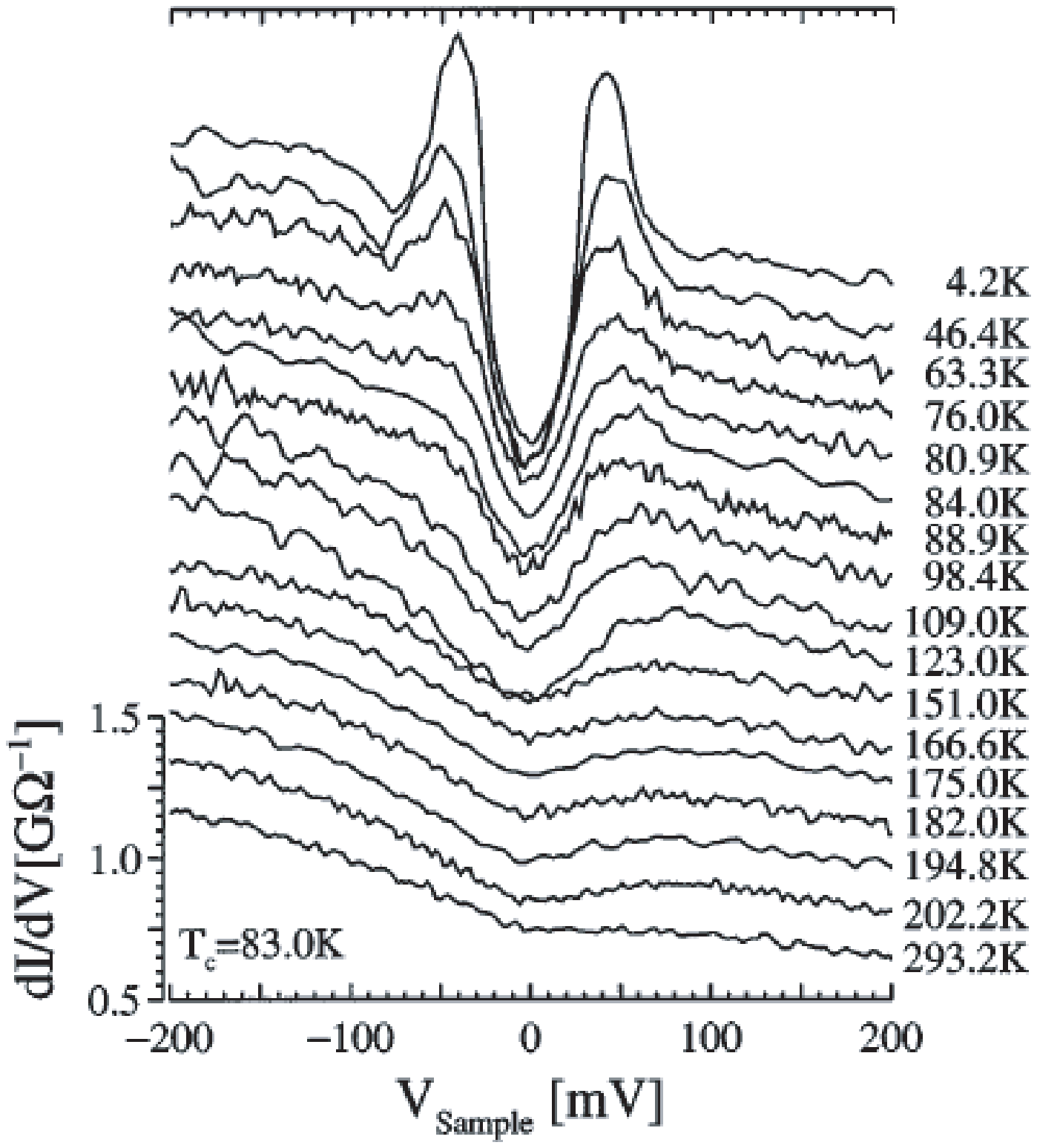}
\caption{Measured density of states at temperatures in the pseudogap phase obtained by Renner et al., Ref. (\onlinecite{renner})in Bi-2212, by scanning tunneling microscopy}
\label{fig. 6}
\end{center}
\end{figure}

   The specific heat coefficient, $\gamma(T)\equiv C_v/T$ and the magnetic susceptibility $\chi(T)$ at different $x$, calculated from the density of states of fig.(5) are presented in fig. (7) and (8) together with the experimental results. The specific heat results in Ref. (\onlinecite{loram}) for various $x$ are rescaled with a $T_p(x)$ to lie as shown in fig.(7) with a common ratio $D_0(x)/T_p(x) = 2.5$. The values of $T_p(x)$ are within the uncertainty the same as the values where the pseudogap is seen in the specific heat or the resistivity. Considering the errors in extracting the electronic specific heat from the measured specific heat, no more sophisticated fitting is warranted, although clearly the fitting could be further improved by choosing a slightly variying $D_o/T_p(x)$ for different $x$.

  The temperature dependence of the measured $\chi(T)$ for a particular $x$ is compared with experiment in fig.(8) with again $D_0/T_p=2.5$. Within the uncertainties in the determination, the experimental $C_v(T,x)/T\chi(T,x)$ is independent of temperature \cite{tallon}, in agreement with the result here. The agreement of the calculated thermodyanmics with the experiments should give one confidence in the underlying theoretical ideas.

    \begin{figure}[htbp]
\begin{center}
\includegraphics[width=0.8\textwidth]{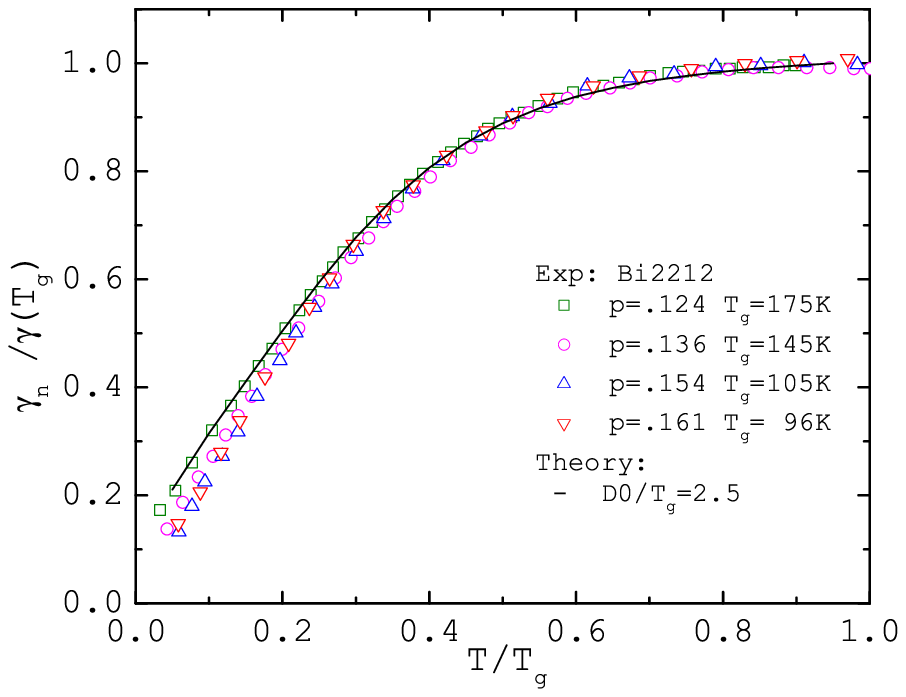}
\caption{Calculated specific heat coefficient $\gamma(T)$ for various  dopings $x$ compared to the values deduced from experiments; The specific heat data was extracted from Fig 12 of J.W. Loram, J. Luo, J.R.
Cooper, W.Y. Liang and J.L. Tallon, J. Phys. Chem. Solids 62, 59 (2001).). \label{fig.6}}
\end{center}
\end{figure}

  \begin{figure}[htbp]
\begin{center}
\includegraphics[width=0.8\textwidth]{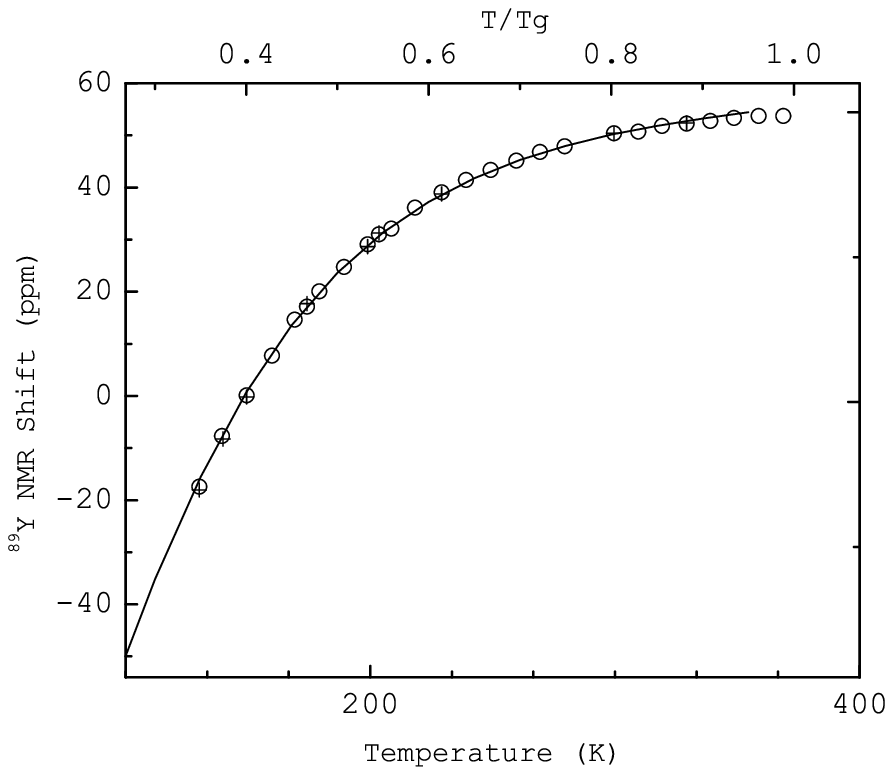}
\caption{Calculated magnetic susceptibility as a function of temperature compared with that deduced from Knight shift measurements reported in Fig 3 of G. V. M. Williams, J. L. Tallon, J. W.
Quilty, H. J. Trodahl, and N. E. Flower, Phys. Rev. Lett. 80, 377 (1998)
. Note that Knight shift measures both the diamagnetic and the paramagnetic susceptibility. In the comparison with the experiments, a temperature independent diamagnetic term has been added to the calculated values.}
\label{fig. 7}
\end{center}
\end{figure}

   From Eq.(72), it follows that the  low temperature specific heat  $C_v$ is
          predicted to be $\sim T^{3/2}$ and magnetic susceptibility $\chi \sim T^{1/2}$ .
        Due to the intervention of superconductivity, it is hard to test these power laws accurately.   One can deduce the
continuation to the T-dependence below $T_c$ of $C_v(T)$ by invoking
conservation of entropy on the available measurements and the $T=0$ limit.
$C_v (T) \sim T^{3/2}$ for $T << T_{x}$ is then not inconsistent
while $C_v (T) \sim T$ clearly is inconsisitent. The temperature dependence of $C_v$ in the pseudogap regime implies that the temperature dependence and magnitude of superfluid density $\rho_s$ in the pseudogap regime are quite different from usual. This is a subject for further investigations.

  \section{Concluding Remarks}

    This work rests on the development of two novel ideas using very elementary methods: Time-reversal violating states in a metal  due to large finite range interactions and that such states necessarily have anisotropic gaps at the chemical potential. It is suggested that these determine the properties of region (II) of the phase diagram ending at the putative QCP, which is the basis for the singular fluctuations responsible for the properties in region (I) and the glue for superconductive pairing.
 The claim that this investigation provides the basic framework for a microscopic theory of the cuprates can only be made if the unusual predictions made are further confirmed and verified. I also discuss in this section, the shortcomings of the mean-field theory in this paper in relation to the properties of the Cuprates.

 \subsection{Predictions}
 Two classes of verifiable predictions have been  made in this and related work:
a) The pseudogap state violates Time-reversal without altering the translational symmetry and ends at a QCP inside the superconducting dome of compositions.
and b) The specific features of the dispersion and the resulting thermodynamic properties in the pseduogap state.

Concerning a), the most dramatic prediction, as has been mentioned, has been verified \cite{kaminski} through specially designed ARPES experiments with polarized photons. The symmetry of the TRV state is consistent with the $\Theta_{II}$ state. An important part of the verified  prediction is that the magnitude of the measured time-reversal violation in a given direction in the Brillouin zone is the same for all states in the conduction band measured \cite{kaminski} (in an energy range of about 0.4 eV below the chemical potential).  According to the theory, these should all be given by the parameter $\theta$.

 These results need to be confirmed independently with other technique, preferably more direct; also more compounds with a range of compositions need to be measured. Experiments using other techniques \cite{dimatteo, ng} have  been proposed to test for time-reversal invariance and the specific point-group symmetry proposed for the current patterns. An especially notable prediction \cite{ng} is the nature of the superconducting state for compositions in which the pseudogap occurs in the normal state. This can be verified through experiments related to the Josephson effect.

Several confirmations of the prediction that the pseudogap state ends at a QCP inside the superconducting dome now exist through tunneling \cite{alff} and transport \cite{greene, boebinger} experiments in a magnetic field.

Concerning b), the principal predictions are
(i) that the dispersion of quasi-particles in the pseudogap state near the gap-edge has the form given by Eq.(70) and exhibited in fig. (3). To check this the spectral function including the self-energy must to be used to deduce the quasi-particle dispersion.
(ii) that the dispersion has point-group symmetry in a single -domain given by Eqs.(86,87  ) of the Appendix.
(iii) that the specific heat and the magnetic susceptibility at very low temperatures in the pseduogap state have the temperature dependence discussed above. The specific heat ( and the related experiment of thermal conductivity ) need to be done in a high enough magnetic field to suppress superconductivity.

A very important issue is the specific heat at the onset of the pseudogap. As discussed below the transition may not have a singularity in the specific heat. But there is no reason why there should not be a bump.  The deduction of the electronic specific heat through subtraction of the lattice heat capacity by comparison between samples for temperatures above $40K$ (where it is less than $1\%$ of the total ) is not accurate enough to decipher bumps. Measurements at much lower temperatures are required which must be done in a large enough field to suppress superconductivity with the transition to the pseudogap preserved.

It is obvious that the fluctuation spectra of the form of Eq.(1) observed in the region (I) of the phase diagram at long wavelengths by Raman scattering should be modified in the classical fluctuation regime qualitatively as in Eq.(29). This is consistent with some recent observations \cite{caprara}. The detailed behavior can only be predicted by a proper theory in the fluctuation regime (see below) and not be a mean-field theory supplemented by Gaussian fluctuations as in this paper. The result derived here that the fluctuations occur in specific IR's given by the coupling of the fluctuation vertex to the Fermions is likely to survive since it is based on symmetry only.

   \subsection{Specific heat near the Transition}

A difficulty in regarding the pseduogap region as a distinct phase is that there is no singularity in the specific heat at the temperature where it begins to be observed in many other experiments. In this connection, it is worthwhile looking at the statistical mechanics for a TRV transition into  the $\Theta_{II}$ phase. This phase
breaks both time-reversal and inversion. A statistical mechanical model
describing the transition to such a phase is that of two-coupled Ising model. A realization of the model is the (anisotropic)
 Ashkin-Teller model:
 \begin{eqnarray}
 H = - J\sum_{<ij>} [(\mu_i \mu_j +x\lambda_i\lambda_j)+ y(\mu_i\mu_j \lambda_i\lambda_j)]
 \end{eqnarray}
 where the sum covers nearest neighbors on a lattice.  For $-1<y<1$ at $x=1$, the
 model in two-dimensions has a line of critical points with variable critical
 exponents.
 We may identify $\langle\mu\rangle$ as the time-reversal order parameter and
  $\langle \lambda \rangle$ as the inversion breaking order parameter.
   $J,x,y $ are positive in our problem signifying {\it{ferromangetic}}
  alignment of $\langle\mu\rangle$, $\langle\lambda\rangle$
   as well as of $\langle\mu\lambda\rangle$. In this case the specific heat exponent
   $\alpha$ varies from 0 to $-2/3$ as $y$ varies from 0 to 1, i.e. the specific heat has no divergence \cite{ashkin-teller}. The anisotropic model ($x \ne 1$) also shares some of these special features.

   The specific heat is part of the general problem of deriving the fluctuation spectra at the TRV/pseudogap transition in the classical regime as well as in the quasi-classical regime, (i.e. Region I in the phase diagram of fig.(1)), which lead to the MFL properties as well as provide the glue for pairing.
 In this connection, it is worth noting that the  Ashkin-Teller model has interesting topological excitations. These may well be relevant to deriving Eq. (1).

       \subsection{Shortcomings}

  Some major blank spaces in the picture have already been mentioned; some others ought to be mentioned. The most important of these is the derivation of the phenomenolgical Eq.(1) and  superconductivity. The fluctuation spectra of Eq. (1) has the right energy scale and coupling constant (both deduced from expriments) to give the right scale of superconducting transition temperatures. The developments in this paper identify the nature of the fluctuations to be time-reversal and chirality fluctuations which condense to give the 'pseudogap' region of the phase diagram.

   The major issue of principle about superconductivity is how a (nearly) momentum independent fluctuation spectra, which predicts the observed momentum independent single-particle self-energy, can lead to d-wave superconductivity. The answer lies in the momentum dependence of the coupling functions of the current fluctuations to the fermions. As noted earlier, the spectra of Eq.(1) acquires a dependence on the internal momentum ${\bf k}$ of the particle-holes to represent the IR of the fluctuation spectra. In a separate paper, I show that this leads to a momentum independent self-energy but a momentum dependent pairing vertex.

    Some other issues worth investigating are the competition between the pseudogap state investigated here and the AFM state near half-filling, the transport properties  in the pseudogap state and the derivation of the unusual superfluid density in the pseudogap state.  A complete solution of the self-consistency equations in section IV-B is also desirable. Many of these  issues which involve detailed calculations are worth the effort only if the principal results of this paper are verified by experiments.

  subsubsection{Acknowledgements}:  I am grateful to Dr. Lijun Zhu for the figures (4,5,6 and 7) and the comparison with experimental data shown there. I have also benefitted enormously from conversations with so large an array of experimentalists and fellow theorists that they cannot all be mentioned here. Special acknowledgements are due to Adam Kaminski, Elihu Abrahams and Peter Woelfle. Part of this work was done at Bell laboratories and part of the mansucript written at Aspen Center for Physics and during visits to Institute for Condensed matter Physics at University of Karlsruhe as a Humboldt prize recipient.

      \section{Appendices}

\subsection{a: Calculation of energies}

To determine the mean-field value of the order parameter, the
dependence of the  energy of the states of the conduction band on
the invariant flux $\Phi(\bf k)\equiv (\theta_x({\bf k}) -\theta_y({\bf k})) $ is calculated. To do this in general , the $3\times 3$ matrix , (17) or (21) needs to be diagonalized. This leads to a rather messy expression. Answers with the correct symmetry may be obtained by calculating the energy perturbatively in $t_{pp}/t_{pd}$. Actually, this parameter may be no smaller than about $1/2$. Treating it as small has the advantage of obtaining results analytically and no disadvantage as far as the qualitative validity of the results is concerned. The change in energy of the conduction band is

\begin{equation}
\delta\epsilon_{c{\bf k}} \approx
8\frac{\bar{t}_{pp}^2}{\epsilon_{c{\bf k}}^0}\frac{(s_x^2s_y^2)^2}{s_{xy}^4}\cos(2(\theta_x({\bf k}) -\theta_y({\bf k})).
\end{equation}
Here  $\epsilon_{c{\bf k}}^0=2{\bar t}_{pd}s_{xy}({\bf k})$ is the difference in energy of the conduction band and the non-bonding band for $t_{pp}=0$. $\theta_{x,y}({\bf k})$ for the $\Theta_I$ and $\Theta_{II}$ phases are given by Eqs. (18) and (19).
$R_{I,II}$ and $\phi_{I,II}$ are to be determined variationally.

We write here the mean-field energy which is dependent on the order parameter up to order $(\frac{R}{\bar{t}_{pd}})^4$.  For the $\Theta_I$ phase
\begin{eqnarray}
\delta F(R_I, \phi_I) \approx
\frac{R_I^2}{V} -16 \sum_{\bf k}\left(1-f(\epsilon_{c{\bf k}})\right)\frac{\bar{t}_{pp}^2}{\epsilon_{c{\bf k}}^0}[\frac{R_I}{2\bar{t}_{pd}})^2(\sin^2\phi_I)- \\ \nonumber  2(\frac{R_I}{2\bar{t}_{pd}})^4(\sin^2\phi_I \cos 2\phi_I)]
\frac{s_x^2s_y^2}{s_{xy}^4}\left(s_x^2s_y^2\right).
\end{eqnarray}

For the $\Theta_{II}$ phase,
\begin{eqnarray}
\delta F(R_{II}, \phi_{II}) \approx \frac{R_{II}^2}{V} -16\sum_{\bf k}\left(1-f(\epsilon_{c{\bf k}})\right)\frac{\bar{t}_{pp}^2}{\epsilon_{c{\bf k}}^0}
[(\frac{R_{II}}{2\bar{t}_{pd}})^2(\sin^2\phi_{II})- \\ \nonumber 2(\frac{R_{II}}{2\bar{t}_{pd}})^4\frac{\left(s_xc_y \pm c_xs_y\right)^2}{s_{xy}^2}(\sin^2\phi_{II}\cos 2\phi_{II})]
\frac{s_x^2s_y^2}{s_{xy}^4} \left(s_xc_y \pm c_xs_y\right)^2.
\end{eqnarray}

Compare the coefficients of the $(R/t_{pd})^2$ terms in Eqs.(80) and
(81). Only the last factors in the parenthesis are different and
the rest of the term is negative at each point
$(k_x,k_y)$. Then it is easy to see
 that the decrease in energy of the state
$\Theta_{II}$ is always lower than of the $\Theta_I$ phase such that for fixed parameters the transition temperature to the former is larger than the latter.
Although this is true in the simplest model interactions, it is
may not be generally true.

On minimizing, $\phi_0$ is  $\pm \pi/2$ for either phase . Also $R_0 \ne 0$ at $T=0$ for the $\Theta_{II}$
phase only if:
\begin{eqnarray}
\frac{\bar{t}_{pd}^2}{V} < \sum_{{\bf k}>{\bf k}_f} \frac{8
\bar{t}_{pp}^2s_x^2(k)s_y^2(k)}{\epsilon_{{\bf k}}^0s_{xy}^4(k)}(s_xc_y \pm s_yc_x)^2
\end{eqnarray}

For $\epsilon_d \ne 0$, $\bar{t}_{pd}^2 s_{xy}^2$ is replaced
approximately by $\bar{t}_1^2 s_{xy}^2 +\epsilon_d^2$.

For $\Theta_{I}$ phase the same condition gives a minimum value of
V larger by a numerical factor of $O(1)$ with the other parameters
fixed.

Note that the $x$ dependence occurs through that of  $\bar{t}_{pd}$ and $\bar{t}_{pp}$ besides the Fermi-energy.  Eq.(82) is satisfied only for $x$, the deviation of doping from hal-filling, less than
a critical doping $|x_c$. This may be estimated from the integral in Eq.(82);
the critical doping at $T=0$ is
\begin{eqnarray}
x_c \approx  V \bar{t}_{pp}^2/(2 \bar{t}_{pd}^3\pm V \bar{t}_{pp}^2).
\end{eqnarray}
Here the plus sign is for the deviation of electron density and minus sign for the deviation of hole density from half-filling.

Similarly the value of the order parameter at $T=0$ may be
estimated from the energy expanded to order $(R_0^4)$ to be
\begin{eqnarray}
\theta_0 \equiv <R>/2\bar{t}_{pd} \approx  x_c|x_c-x|^{1/2}
\end{eqnarray}

The transition temperature $T_c$ for a fixed $x$ can be estimated
from Eqs.(80,81). The condition obtained can be written in terms of $|x_c-x|$ and is given in Eq. (24).

The numerical coefficents in all the equations above are quite approximately determined. The
aim here is to show only their dependence on the parameters of the
model.

 \subsection{Energy of Fluctuations}

The energy of uniform fluctuations of $\phi$ about $\phi_0 = \pm
\pi/2$ can be calculated from Eq.(80,81) and is given, for example in
the $\Theta_{I}$ phase by
\begin{eqnarray}
\Omega_{0, \phi} \approx t^2_{pp}\theta_1^2\sum_{{\bf
k}}8(\epsilon_{c{\bf k}}^0)^{-1}\frac{s_x^4s_y^4}{s_{xy}^4},
\end{eqnarray}
which is of $O(\theta_0^2\frac{t^2_{pp}}{t_{pd}})$.

In a given domain of the $\Theta_{II}$ phase, with reflection
symmetry broken about either $k_x = 0$ or $k_y = 0$ planes, the
characteristic  energy of oscillations to the other domain may
also be estimated similarly. In Sec.(IV), I have called such
fluctuations the mixing fluctuations. Their energy at long wavelengths can be
calculated by allowing the order parameter to be a mixture of the
two domains in the mean-field calculations and picking up the
coefficient of the quadratic term $(\delta \theta_m)^2$ in the mixing:
\begin{eqnarray}
\sin \Phi({\bf k}) = \Phi_0 (c_x/s_x+c_y/s_y) +\delta\theta_m(c_x/s_x-c_y/s_y).
\end{eqnarray}
The fluctuation energy proportional to $(\delta\theta_m)^2$ is
\begin{eqnarray}
\Omega_{0,m} \approx\bar{t}^2_{pp}\theta_2^2\sum_{{\bf
k}}8(\epsilon_{c{\bf k}}^0)^{-1}\frac{s_x^2s_y^2}{s_{xy}^4} \left(sin^2((k_x-k_y)a/2)
\cos^2((k_x+k_y)a)\right)
\end{eqnarray}
This  fluctuation energy is seen to be smaller smaller than that
of the mode of the phase of Time-reversal, $\delta \phi$ by about
a factor of 2. The fluctuation energy for the amplitude mode can
be shown to be the largest of the frequencies of the three modes.


\begin{thebibliography}{99}
\bibitem{expts}
See for instance the proceedings of the tri-annual conferences on Materials and Mechanisms of Superconductivity, the latest of which is
\bibitem{errors}
Several calculaional errors in Ref.(\onlinecite{cmv2}) are corrected here.
\bibitem{footnote1}
The Raman spectra measures the response functions as a function of frequency at long wavelengths of  second rank tensors which transform as the IR's of the point-group opf the lattice; A1g, A2g, B1g, B2g etc. for a square lattice. I refer to the time-reversal breaking operators in such representations also as currents. The term is more commonly used only for the odd-symmetry channels which are measured in conductivity experiments.
\bibitem{footnote 2}
This is of-course not uniquely determined by the experiment. One could, for example, envisage a stress-tensor component fulfilling these requirements, but it would then couple linearly to elastic constants giving thereby an anomaly in a transverse velocity or in an optical mode frequency. No such anomalies have been reported.
\end{thebibliography}
   \end{document}